\documentclass[fleqn,usenatbib]{mnras}

\usepackage[T1]{fontenc}

\DeclareRobustCommand{\VAN}[3]{#2}
\let\VANthebibliography\thebibliography
\def\thebibliography{\DeclareRobustCommand{\VAN}[3]{##3}\VANthebibliography}

\usepackage{graphicx}	
\usepackage{amsmath}	
\usepackage{amssymb}	
\usepackage{txfonts}
\usepackage{natbib}
\usepackage{longtable}
\usepackage{multirow}
\usepackage{supertabular}
\usepackage{url}
\usepackage{hyperref}

\title[Complexity of the Bulge MDF]{How Many Components? Quantifying the Complexity of the Metallicity Distribution in the Milky Way Bulge with APOGEE}

\author[A. Rojas-Arriagada et al.]{Alvaro Rojas-Arriagada,$^{1}$ $^{2}$\thanks{E-mail: arojas@astro.puc.cl}
Gail Zasowski,$^{3}$
Mathias Schultheis,$^{4}$
Manuela Zoccali,$^{1}$ $^{2}$
\newauthor
Sten Hasselquist,$^{3}$\thanks{NSF Astronomy and Astrophysics Postdoctoral Fellow.}
Cristina Chiappini,$^{5}$
Roger E. Cohen,$^{6}$
Katia Cunha,$^{7}$ $^{8}$
Jos\'e G. Fern\'andez-Trincado,$^{9}$
\newauthor
Francesca Fragkoudi,$^{10}$
D. A. Garc\'ia-Hern\'andez,$^{11}$ $^{12}$
Doug Geisler,$^{13}$ $^{14}$ $^{15}$
Felipe Gran,$^{1}$ $^{2}$ $^{16}$
\newauthor
Jianhui Lian,$^{3}$
Steven Majewski,$^{17}$
Dante Minniti,$^{18}$ $^{19}$
Antonela Monachesi,$^{14}$ $^{15}$
Christian Nitschelm,$^{20}$
\newauthor
Anna B. A. Queiroz$^{5}$
\\
$^{1}$Instituto de Astrof\'{i}sica, Facultad de F\'{i}sica, Pontificia Universidad Cat\'{o}lica de Chile, Av. Vicu\~{n}a Mackenna 4860, Santiago 8970117, Chile\\
$^{2}$Millennium Institute of Astrophysics, Av. Vicu\~{n}a Mackenna 4860, 7820436, Macul, Santiago, Chile\\
$^{3}$Department of Physics \& Astronomy, University of Utah, Salt Lake City, UT 84105, USA\\
$^{4}$Universit\'e C\^ote d'Azur, Observatoire de la C\^ote d'Azur, CNRS, Laboratoire Lagrange, Blvd de l'Observatoire, F-06304 Nice, France\\
$^{5}$Leibniz-Institut fur Astrophysik Potsdam (AIP), An der Sternwarte 16, D-14482 Potsdam, Germany\\
$^{6}$Space Telescope Science Institute, 3700 San Martin Drive, Baltimore, MD 21218, USA\\
$^{7}$University of Arizona, Department of Astronomy and Steward Observatory, Tucson, AZ 85719, USA\\
$^{8}$Observat\'orio Nacional, Sao Crist\'ovao, Rio de Janeiro 20921400, Brazil\\
$^{9}$Instituto de Astronom\'ia y Ciencias Planetarias de Atacama, Universidad de Atacama, Copayapu 485, Copiap\'o 1531772, Chile\\
$^{10}$Max-Planck-Institut f\"ur Astrophysik, Karl-Schwarzschild-Str. 1, D-85748 Garching, Germany\\
$^{11}$Instituto de Astrof\'isica de Canarias (IAC), E-38205 La Laguna, Tenerife, Spain\\
$^{12}$Departamento de Astrofísica, Universidad de La Laguna (ULL), E-38206 La Laguna, Tenerife, Spain\\
$^{13}$Departamento de Astronom\'ia, Universidad de Concepci\'on, Avenida Esteban Iturra s/n, Casilla 160-C, Chile\\
$^{14}$Instituto de Investigaci\'on Multidisciplinario en Ciencia y Tecnolog\'ia, Universidad de La
Serena, Avenida Ra\'ul Bitr\'an s/n, La Serena, Chile\\
$^{15}$Departamento de Astronom\'ia, Facultad de Ciencias, Universidad de La Serena. Av.
Juan Cisternas 1200, La Serena 1720236, Chile\\
$^{16}$European Southern Observatory, Alonso de Cordova 3107, Vitacura, Casilla 19001, Santiago, Chile\\
$^{17}$Department of Astronomy, University of Virginia, Charlottesville, VA 22904-4325, USA\\
$^{18}$Departamento de Ciencias Fisicas, Facultad de Ciencias Exactas, Universidad Andres Bello, Fernandez Concha 700, Las Condes, Santiago 7591538, Chile\\
$^{19}$Vatican Observatory, V00120 Vatican City State, Italy\\
$^{20}$Centro de Astronom\'ia (CITEVA), Universidad de Antofagasta, Avenida Angamos
601, Antofagasta 1270300, Chile
}

\date{Accepted for publication manuscript. Accepted 2020 September 9.}

\pubyear{2020}

\begin{document}
\label{firstpage}
\pagerange{\pageref{firstpage}--\pageref{lastpage}}
\maketitle

\begin{abstract}
We use data of $\sim$13,000 stars from the SDSS/APOGEE survey to study the shape of the bulge MDF within the region $|\ell|\leq11^\circ$ and $|b|\leq13^\circ$, and spatially constrained to ${\rm R_{GC}\leq3.5}$~kpc. We apply Gaussian Mixture Modeling and Non-negative Matrix Factorization decomposition techniques to identify the optimal number and the properties of MDF components. We find the shape and spatial variations of the MDF (at ${\rm [Fe/H]\geq-1}$~dex) are well represented as a smoothly varying contribution of three overlapping components located at [Fe/H]~=+$0.32$, $-0.17$ and $-0.66$~dex. The bimodal MDF found in previous studies is in agreement with our trimodal assessment once the limitations in sample size and individual measurement errors are taken into account. The shape of the MDF and its correlations with kinematics reveal different spatial distributions and kinematical structure for the three components co-existing in the bulge region. We confirm the consensus physical interpretation of metal-rich stars as associated with the secularly evolved disk into a boxy/peanut X-shape bar. On the other hand, metal-intermediate stars could be the product of in-situ formation at high redshift in a gas-rich environment characterized by violent and fast star formation. This interpretation would help to link a present-day structure with those observed in formation in the center of high redshift galaxies. Finally, metal-poor stars may correspond to the metal-rich tail of the population sampled at lower metallicity from the study of RR Lyrae stars. Conversely, they could be associated with the metal-poor tail of the early thick disc.
\end{abstract}




\section{Introduction}
\label{sec:introduction}
Over the past decades, the bulge of the Milky Way (MW) has been an object of intense study and debate. A vast quantity of observations has contributed to an increasingly high-resolution, detailed characterization of its observed properties and their internal correlations. At the same time, considerable effort has been devoted to producing numerical simulations to provide theoretical prescriptions that link these observed properties with different mechanisms of bulge formation and evolution. 

These efforts demonstrate recognition that the Galactic bulge is a key ingredient in our quest to understand galaxy formation and evolution. It represents the closest example of a mature bulge, which makes possible the detailed study of its resolved stellar content by means of photometric and spectroscopic measurements. The bulge provides an unique opportunity to understand the complex physics of baryons involved in galaxy formation, but at the same time represents a challenge since it requires dense coverage of at least 400~deg$^2$ of heavily extincted sky. 

The two broadest scenarios often invoked to explain bulge formation are motivated by observations of local external galaxies, and aim to account for their morphological dichotomy into the so-called \textit{classical} and \textit{pseudo-bulges} \citep{Kormendy04}. Classical bulges are thought to be the product of mergers of primordial structures that yield spheroidal, massive (relative to the disc), kinematically hot, and isotropic structures, in a $\Lambda-$CDM context \citep{Baugh96,Abadi03}. On the other hand, pseudo-bulges are the end product of the the secular internal evolution of the disc, which rearranges angular momentum and stars to the center and forms a bar, which subsequently undergoes vertical instabilities into a X-shaped boxy/peanut (B/P) bulge \citep{Combes90}. In both cases, the mechanisms are dissipationless processes, driven by stellar dynamics (i.e., not gas dynamics) which assemble central structures from stars that have been previously formed elsewhere.

An additional, complementary scenario of bulge formation emerges from observational evidence at high redshift. Enhanced central star formation or even fully formed bulges have been observed in galaxies at $z\sim2$, which otherwise are still vigorously forming stars in their massive, gas-rich discs \citep{Tacchella15,Nelson16}. This bulge assembly therefore takes place at an epoch preceding the formation of the bar, which for MW-mass galaxies happens at $z\sim1$ \citep[$\sim$8 Gyr ago;][]{Sheth08,Kraljic12,Fragkoudi20}. Characterized by a period of high star formation rate (SFR), the mechanism responsible here seems to produce bulges from the compaction of gas in a dissipative process, which is fast and forms stars in-situ.

It is well-established that the Galactic bulge hosts a bar and has a B/P morphology \citep[e.g.,][]{Weiland94,Dwek95,wegg13}. N-body simulations indicate that such a structure is evidence of an origin in secularly evolved disc \citep{combes_sanders:81,Raha91,Athanassoula05}. This scenario has been reinforced and refined by the results of high resolution hydrodynamical simulations, which in addition to tracking the dynamical evolution of the particles, account for the detailed physics of a large number of baryonic processes \citep{Gargiulo19,Fragkoudi20}.
Additional complexity  has emerged as the outcome of a progressively larger amount of small programs and surveys mapping the stellar content of the inner Galaxy. In particular, it has become evident that this region hosts a variety of stellar populations or components, some of which have distinct kinematical or spatial distributions (e. g. Mira stars; \citeauthor{Catchpole16} \citeyear{Catchpole16}, RC stars; \citeauthor{Zoccali17} \citeyear{Zoccali17}, RR Lyrae; \citeauthor{Kunder20} \citeyear{Kunder20}).

During the past twenty years, enormous effort has gone into obtaining large samples of stellar spectra in the Galactic bulge with the aim of understanding the history of this massive component \citep[for a review of these surveys, see][]{Barbuy18}. However, differences in target selection, stellar tracers, sightlines, and analysis methods long prevented a consistent picture of the bulge.

Starting with the pioneering low resolution spectroscopic study of \citet{Rich88} in Baade's Window, the metallicity distribution function (MDF), and in general, the chemistry of K and M bulge giant stars started to be the object of dedicated observational efforts. Low resolution studies of the order of hundred stars \citep{Sadler96,Minniti96b,Ramirez00} were complemented with high resolution studies of the order of tens of stars, not only in Baade's Window but also down to the high reddening regions close to the midplane, as near-infrared (NIR) spectroscopy was used \citep{McWilliam94,Fulbright06,Cunha06,Cunha07,Rich07}.

The first homogeneous assessment of the MDF in several fields is that of \citet{zoccali08}, who studied the MDF along the bulge's minor axis (between b\,=\,$-4^\circ$ and b\,=\,$-12^\circ$) by high resolution observations of hundreds of stars per field, and confirmed the vertical metallicity gradient suggested earlier by \citet{minniti:95}; this was interpreted as a signature of classical bulge formation. Using a larger sample of red clump stars \citet{Hill11}, revealed for the first time a bimodality in the MDF, the peaks of which are correlated with different kinematical signatures \citep{Babusiaux10}. The metal-rich stars showed kinematics consistent with a bar-driven component, while the metal-poor ones had more isotropic kinematics that were associated with a classical spheroid. A similar metallicity bimodality was suggested from the analysis of the first available sample of dwarf, turn-off and subgiant bulge stars, observed during microlensing events \citep{Bensby11}. In addition, this work suggested a different age-metallicity relation for metal-poor and metal-rich stars: while the former are generally old ($\sim11$~Gyr), the latter span a wide range of ages, with a significant fraction being younger than 9~Gyr. Together with the aforementioned different kinematic behavior, these findings started to reveal an overall complex picture of the bulge stellar content.

Several subsequent studies, using different spectroscopic surveys of the bulge region, confirmed this bimodality and the relationship between metallicity and kinematics:  e.g., Gaia-ESO Survey (GES; \citeauthor{Rojas-Arriagada14} \citeyear{Rojas-Arriagada14}, \citeyear{Rojas-Arriagada17}), GIRAFFE Inner Bulge Survey (GIBS; \citeauthor{Gonzalez15} \citeyear{Gonzalez15}, \citeauthor{Zoccali17} \citeyear{Zoccali17}), and APOGEE (\citeauthor{Ness16} \citeyear{Ness16}, \citeauthor{Zasowski16} \citeyear{Zasowski16},  \citeauthor{Schultheis17} \citeyear{Schultheis17}, \citeauthor{Fragkoudi18} \citeyear{Fragkoudi18}, \citeauthor{Queiroz_2020_starhorse} \citeyear{Queiroz_2020_starhorse}).

Using more than 10,000 stars from ARGOS, \citet{Ness13} found that the MDF can be decomposed into up to five metallicity components, with three of them accounting for the majority of stars. Using the APOGEE DR12 data, \citet{garciaperez18} suggested the presence of four metallicity components which are of different strength. However, their sample did not include the coolest stars with temperatures below 3600\,K. This complexity in the MDF structure, seen from the study of bulge giant stars, was reinforced by the results of the progressively larger sample of microlensed dwarf stars analyzed in \citet{Bensby13} and \citet{Bensby17}. Up to five peaks were identified in their MDF, with the metallicity locations being consistent with those detected in the MDF assembled from the larger ARGOS sample.

In this paper, we use the combined data of APOGEE-1 and APOGEE-2 (Sect.~\ref{sec:data}) to study the bulge's MDF, its statistical properties, and correlations between stellar metallicity and kinematics over a large area of sky, including the still poorly explored inner degrees of the Galactic plane. 

The structure of the paper is as follows. In Sect.~\ref{sec:data} we describe the adopted dataset and the computed set of spectro-photometric distances. We explore potential sources of bias in the MDF, as well as the appropriateness of the often adopted radial limit (${\rm R_{GC}\leq3.5~kpc}$) used to select samples of bulge stars in Sect.~\ref{sec:define_likely_blg_smpl}. The shape of the bulge MDF and its spatial variations are explored in Sect.~\ref{sec:mdf}, where it is parametrized with a Gaussian Mixture Model analysis as well as a Non-negative Matrix Factorization decomposition. Finally, in Sect.~\ref{sec:discussion} we discuss our results in the context of previous spectroscopic surveys of the bulge, as well as in the more general context of galaxy formation.

\section{Data}
\label{sec:data}

\subsection{APOGEE and Gaia}
\label{subsec:apo_gaia_data}

We use fundamental stellar parameters effective temperature ($T_{\rm eff}$), surface gravity ($\log{g}$) and metallicity ([M/H]\footnote{Throughout this paper we use [M/H] as the estimate of metallicity, which results from the full-spectrum fitting performed by ASPCAP.  These values are consistent with the [Fe/H] measurements, so we compare our [M/H] MDF components directly with the [Fe/H] of the literature components.}) from the Apache Point Observatory Galactic Evolution Experiment \citep[APOGEE;][]{majewski:17}. APOGEE is a high-resolution, near-infrared (NIR) spectroscopic survey designed to perform far-reaching chemical cartography of the Milky Way stellar populations, using hundreds of thousand of stars. The main targets of the survey are giant stars (RGB, AGB, and RC), which are intrinsically luminous tracers present in nearly all stellar populations \citep{zasowski2013,Zasowski_2017_apogee2targeting}. By observing at NIR wavelengths, APOGEE overcomes much of the extinction imposed by the large amount of dust present in the Galactic plane, especially towards the Galactic bulge, which has limited past observational efforts in this important region of the Galaxy. 

APOGEE, a component of both SDSS-III and -IV \citep{Eisenstein_11_sdss3overview,Blanton_2017_sdss4}, observes in the NIR $H$-band ($1.51-1.70$~$\mu$m) using two custom-built, high-resolution ($R \sim 22,500$) spectrographs at Apache Point Observatory's 2.5~m Sloan Telescope and Las Campanas Observatory's 2.5~m Ir\'en\'ee du~Pont telescope \citep{Bowen_1973_duPontTelescope,gunn2006,Wilson_2019_apogeespectrographs}. APOGEE spectra are extracted, wavelength calibrated, and radial velocities (RVs) are computed using the pipeline described in \citet{Nidever_2015_apogeereduction}; stellar fundamental parameters and abundances of up to 26 elements (including alpha, iron-peak, odd-Z and neutron-capture elements) are computed using the APOGEE Stellar Parameters and Chemical Abundances Pipeline \citep[ASPCAP;][]{garcia16}. An overview of the APOGEE parameter calibrations, data products, and elemental abundances can be found in \citet{Holtzman_2018_dr13dr14apogee} and \citet{Jonsson_2018_dr13dr14abundances}.

The base catalog for our sample selection has been reduced and analyzed either with the APOGEE pipeline version used in SDSS Data Release 16 \citep[DR16;][]{Ahumada20,Jonsson20}, or a very similar one with a slightly updated data reduction version (r13). The full catalog includes $\sim$134,500 additional stars observed after those released in DR16 (through November 2019), and additional visits for $\sim$56,000 others.

We adopt proper motions for our sample from {\it Gaia}~DR2 \citep{GaiaCollab_2018_gaiaDR2}. In addition, we adopt the renormalized unit weight error (RUWE) values for all the stars in our sample from the official release as available from the Gaia Archive. The RUWE is a recommended astrometric-quality diagnostic which can be used as a criterion to select good astrometric solutions. We do not use these values in selecting the bulge sample, but we include them in our study of the spatial variation of global kinematical properties of the sample (Sect.~\ref{subsec:35kpc_justification}).
On the other hand, we do not use {\it Gaia} parallaxes as they become error dominated for stars beyond $3-4$~kpc, making them unsuitable to estimate distances for stars in the inner Galaxy. Instead, we compute spectro-photometric distances as described in the next section.

\subsection{Distances and orbits}
\label{subsec:dists_orbits}
We calculated spectro-photometric distances for the whole set of stars available in the interim APOGEE catalog (so, including data beyond the public DR16). To this end, we incorporated the stellar properties ${\rm T_{eff}}$, $\log(g)$, [M/H], and the 2MASS {\it JHK$_s$} photometry into the spectro-photometric method described in \citet{Rojas-Arriagada17} and \citet{Rojas-Arriagada19}.

In summary, we compare a large set of theoretical isochrone points to the parameters $T_{\rm eff}$, $\log{g}$, and [Fe/H] of each star, and compute the distances from each star to each isochrone point in the theoretical space. These parameter-space distances serve as weights from which the most likely theoretical physical properties (e. g. the luminosity) of the star can be computed. A number of extra multiplicative weights are defined to account for the evolutionary speed of the points along the isochrones and for the IMF. Using these weights, the most likely absolute magnitudes ($M_J,\ M_H,\ M_{K_s}$) of the observed stars can be computed as the weighted mean or median of the theoretical values of the whole set of isochrone points. The computed absolute magnitudes are then compared to the observed photometry, allowing us to estimate the line-of-sight reddening and distance modulus. No prior on the Galactic stellar density is used.
For these computations, we adopted a set of PARSEC\footnote{Available at \url{http://stev.oapd.inaf.it/cgi-bin/cmd}} isochrones \citep[version 3.1, ][]{Bressan_2012_parsec,Marigo_2017_parseccolibri}, spanning ages from 1 to 13~Gyr in steps of 1~Gyr, and metallicities from $-2.2$ to +$0.5$ in steps of 0.1~dex.

In Appendix~\ref{ap:sec:distance_validation}, we validate our distances against other established metrics: the {\sc StarHorse} \citep{Queiroz18,Queiroz_2020_starhorse} and {\sc astroNN} \citep{Leung19} pipelines, {\it Gaia} Bayesian distances \citep{Bailer-Jones18}, open clusters, and the distances to the Large Magellanic Cloud and Sagittarius dwarf galaxy. Overall, we find an approximate conservative uncertainty of $\sim$25\% for the typical RGB stars that dominate our sample.

For the rest of the paper, we use Galactocentric cylindrical distances ($R_{\rm GC}$) computed from our spectro-photometric heliocentric distances and the stellar $(l,b)$ coordinates, assuming $R_\odot$\,=\,8.2\,kpc \citep{Bland-Hawthorn16}. We also adopt the Galactocentric velocity $V_{\rm GC}$ for our analysis, that is, the heliocentric radial velocity corrected for the solar reflex motion.

The combination of spectro-photometric distances, \textit{Gaia} proper motions, and APOGEE radial velocities allow us to estimate orbital parameters by integrating orbits under a prescription of the Galactic potential. We use the {\sc galpy}\footnote{Available at \url{http://github.com/jobovy/galpy}} code \citep{Bovy15}, adopting the \texttt{MWPotential2014} model for the Milky Way gravitational potential. This model is a superposition of a Hernquist bulge \citep{Hernquist90}, a Miyamoto-Nagai disc \citep{Miyamoto75}, and a Navarro-Frenk-White halo \citep{Navarro97}, which contribute 5\%, 60\% and 35\% of the rotational support at the solar circle, respectively.

From the full phase-space information available for each star ($\alpha$, $\delta$, $\mu_\alpha\cos(\delta)$, $\mu_\delta$, $V_{\rm GC}$, $d$), the stellar orbits are integrated over 10~Gyr. In the calculations, the in-plane distance of the Sun from the Galactic center is adopted as $R_\odot$\,=\,8.2\,kpc, the velocity of the Local Standard of Rest (LSR) as $V_{\rm LSR}$\,=\,220\,${\rm km\ s^{-1}}$ \citep[see][]{Bovy12a}, and the peculiar velocity of the sun respect to the LSR as ($U,V,W$)$_\odot$\,=\,(11.1, 12.24, 7.25)\,${\rm\ km\ s^{-1}}$  \citep{Schoenrich10}.
We estimate uncertainties for all computed orbital parameters by generating 600 random Gaussian realizations of the set of observed parameters from their respective individual uncertainties. From these variations of the initial conditions, we obtain distributions of the orbital parameters from which $1\sigma$ errors are estimated; these are used to restrict the sample solely to demonstrate the $R_{\rm GC}$ limit in Sect.~\ref{subsec:35kpc_justification}.

\subsection{Selection of a clean inner disc/bulge sample}
\label{subsec:sel_gen_clean_sample}
Before examining the bulge MDF, we construct a clean sample of inner MW APOGEE stars with high-quality, reliable fundamental parameters but with no distance restrictions.  We use this sample in Sect.~\ref{sec:define_likely_blg_smpl} to determine the optimal $R_{\rm GC}$ limit for a bulge selection.

To select this clean inner sample, we start with the interim {\tt allStar} file of all APOGEE data observed through November 2019 (MJD~58814) and consider stars with $|\ell| \le 16^\circ$ and $|b| \le 15^\circ$.  
To ensure reliable ASPCAP values for the parameters of interest, we require a minimum signal-to-noise ratio (SNR) of 60 and that the following STARFLAG\footnote{\url{https://www.sdss.org/dr16/algorithms/bitmasks/\#collapseAPOGEE_STARFLAG}} bits be set to zero: BAD\_PIXELS, VERY\_BRIGHT\_NEIGHBOR, and PERSIST\_HIGH. We require the NO\_ASPCAP\_RESULT and the STAR\_BAD bits of the ASPCAPFLAG\footnote{\url{https://www.sdss.org/dr16/algorithms/bitmasks/\#collapseAPOGEE_ASPCAPFLAG}} bitmask to be zero, and we remove stars with visit-to-visit radial velocity variations (VSCATTER) greater than 0.6~km~s$^{-1}$. The selected SNR limit includes stars below the APOGEE goal of ${\rm SNR\geq80}$ per pixel, but is set conservatively above the SNR\,=\,50 limit over which ASPCAP provides reliable parametrizations \citep{garcia16}. The VSCATTER limit is set to remove a minor fraction of stars in the tail of the VSCATTER distribution, dominated by binary systems or stars with some other problem affecting their RV measurements. In addition to the previous cuts, we remove any stars that pass them but do not have valid ASPCAP values and/or spectro-photometric distances.

\begin{figure*}
\begin{center}
\includegraphics[width=0.9\textwidth]{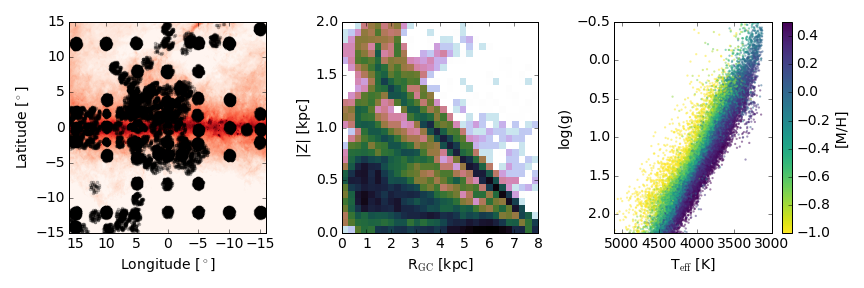}
\caption{
The clean general inner Galaxy sample as defined in Sect.~\ref{subsec:sel_gen_clean_sample}. 
{\it Left:} Galactic ($l,b$) of the stars (black points) against the \citet{Schlegel_1998_dustmap} E(B-V) map (in red). 
{\it Center:} Galactocentric radius ($R_{\rm GC}$) and height from the midplane of the stars.
{\it Right:} Kiel diagram, colored by [M/H].
}
\label{fig:sample}
\end{center}
\end{figure*}

To eliminate biases due to non-standard target selection, we remove any star not flagged as a main survey target using the EXTRATARG\footnote{\url{https://www.sdss.org/dr16/algorithms/bitmasks/\#collapseAPOGEE_EXTRATARG}} bitmask. This leaves several inner Galaxy stars that were chosen as part of smaller APOGEE subprograms (in addition to the main survey); after evaluating those programs' target selection and the resulting samples, we remove stars whose PROGRAMNAME in the {\tt allStar} file is set to {\it cluster\_gc}, {\it clusters\_gc1}, {\it geisler\_18a}, {\it geisler\_19a}, {\it sgr}, or {\it sgr\_tidal}. The resulting sample after all these selections comprises 27,806 stars and is shown in Figure~\ref{fig:sample}.

\section{Definition and bias corrections of our bulge sample}
\label{sec:define_likely_blg_smpl}
The objective of this work is to study the shape of the MDF as an observational proxy of the complex mix of structures assembling the stellar content of the inner Milky Way. In this context, there are two main issues we examine before defining a selection of bulge stars: potential sources of sampling biases in the APOGEE data (Sect.~\ref{sec:bias_corrections}) and the most appropriate cylindrical distance cut to define the bulge spatial region (Sect.~\ref{subsec:35kpc_justification}).
We incorporate our findings into the definition of our bulge sample in Sect.~\ref{subsec:sel_likely_bulge}.

\subsection{Sample bias corrections}
\label{sec:bias_corrections}
As we want to use the APOGEE spectroscopic data to study the underlying structure of the bulge MDF and its spatial variations, we must therefore account for potential sources of sampling biases.
We calculated selection corrections for two different types of biases in our sample: $i$) observational bias due to stars of different metallicities having different probabilities of being observed by APOGEE ($P_{\rm obs}$), and $ii$) analysis bias due to stars with some intrinsic metallicity--$T_{\rm eff}$ combinations not having reliable ASPCAP metallicities ($P_{\rm[M/H]}$).

\begin{figure*}
\begin{center}
\includegraphics[angle=0,trim=0in 0in 0in 0in, clip, width=0.95\textwidth]{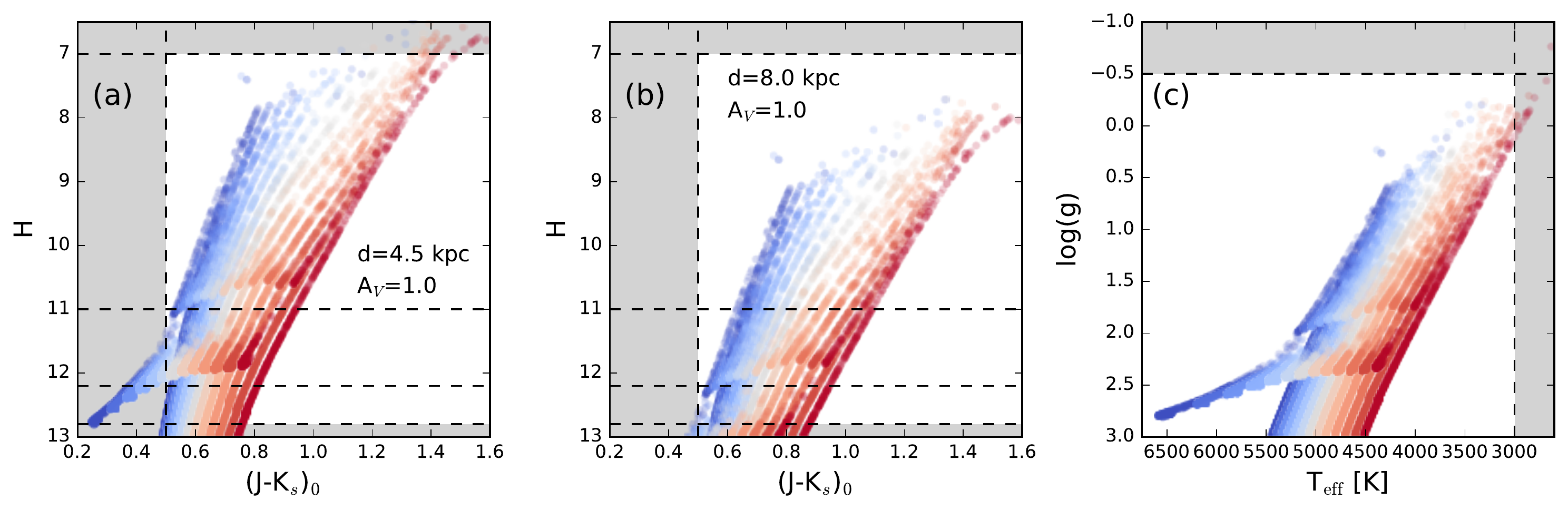} 
\caption{Example of sample selection limits, using 10~Gyr SSPs (based on MIST isochrones) with [Fe/H] between $-2$ (blue) and $+0.4$ (red). Panel (a) simulates the photometric properties of populations with a heliocentric distance of 4.5~kpc ($R_{\rm GC} \approx 3.5$~kpc) and an extinction of $A_V=1$~mag. Dashed lines indicate the color and magnitude limits used by APOGEE, with the shaded regions excluded from the main survey sample. Panel (b) is similar to (a), but for these same populations shifted to the approximate distance of the Galactic Center. Panel (c) shows the $T_{\rm eff}$--$\log{g}$ distribution of these stars, with the limits of the ASPCAP parameter grid shown as dashed lines.}
\label{fig:limits}
\end{center}
\end{figure*}

\begin{figure*}
\begin{center}
\includegraphics[angle=0,trim=0in 0in 0in 0in, clip, width=0.6\textwidth]{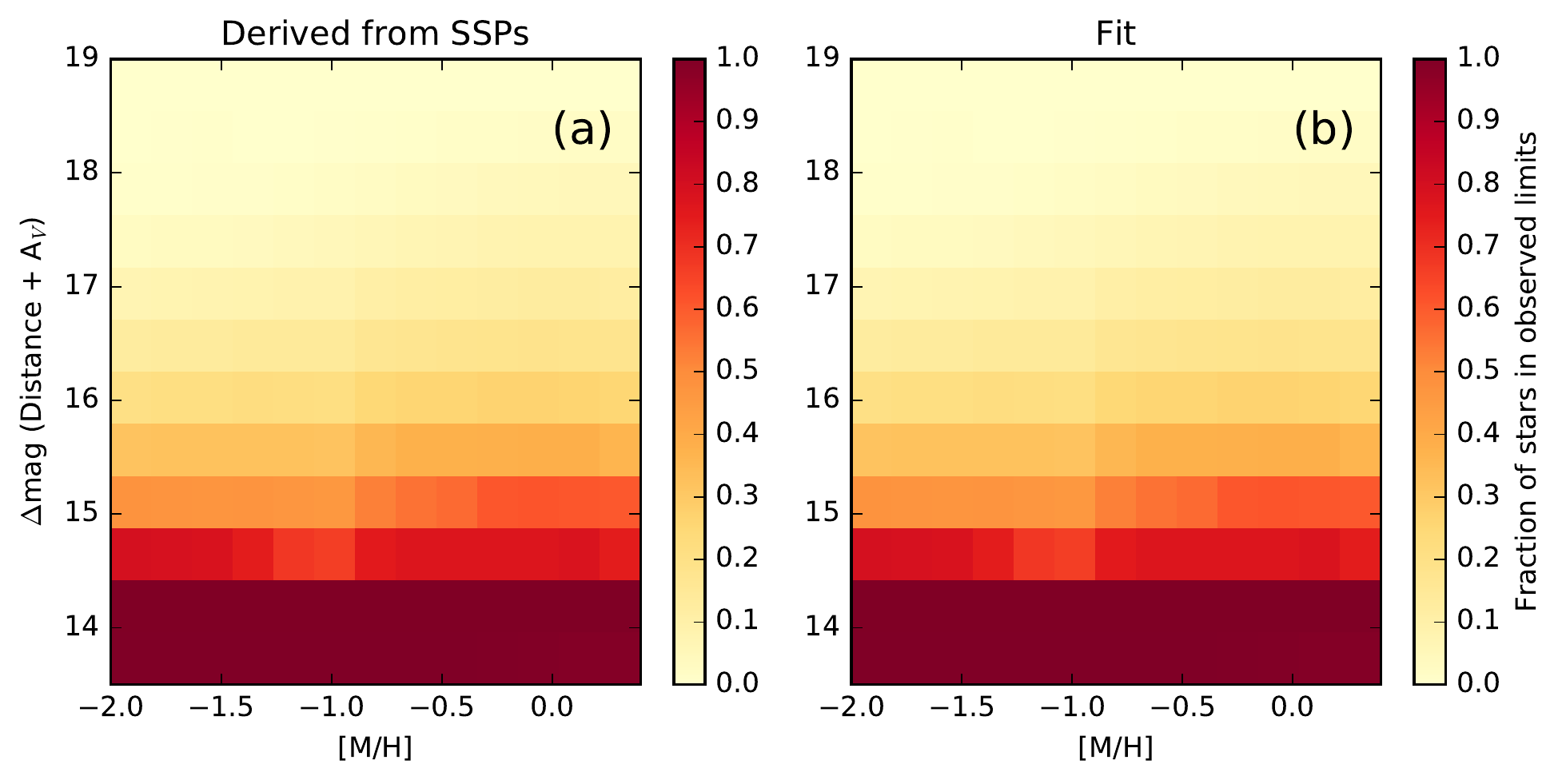} 
\includegraphics[angle=0,trim=0in 0in 0in 0in, clip, width=0.3\textwidth]{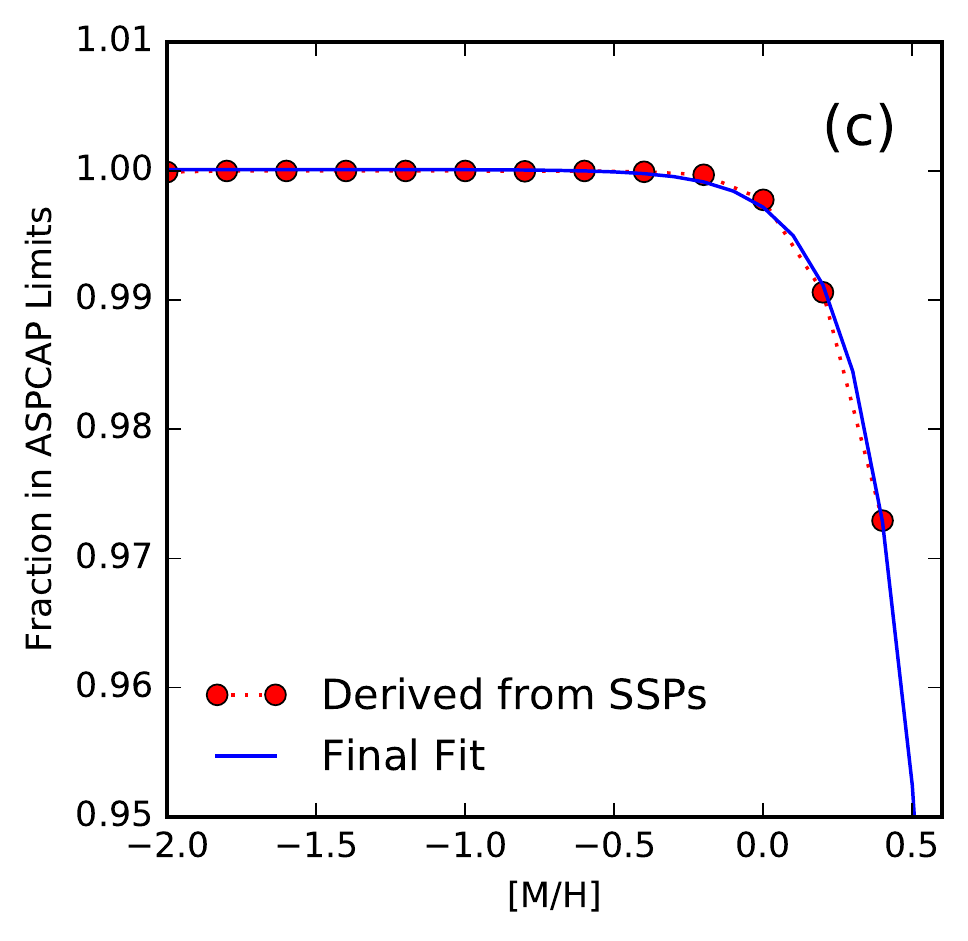} 
\caption{Panels (a)--(b): Example of fit to the ``observed'' fraction of the RGB+AGB branch, as a function of stellar [M/H] and $\mu$ for the selection bin defined by $(J-K_s)_0 \ge 0.5$ and $7 < H < 12.2$ (see Sect.~\ref{sec:bias_corrections}). Panel (c): Fit to the fraction of simulated RGB+AGB stars within ASPCAP's analysis range.}
\label{fig:bias_fit}
\end{center}
\end{figure*}

For the first of these -- the observational probability -- we simulated observing simple stellar populations (SSPs) as a function of [Fe/H], distance, and extinction (see examples in Figures~\ref{fig:limits}a and \ref{fig:limits}b). We generated SSPs based on 10~Gyr MIST isochrones \citep{Paxton11,Dotter16} for $\rm -2.0 \le [Fe/H] \le +0.4$, with $\Delta$[Fe/H]~=~0.2~dex, using a \citet{Kroupa_2001_imf} IMF. We computed $H$-band photometry for each SSP at a range of distances ($d$) and extinctions ($A_H$), together as $\mu = 5 \log{d} - 5 + A_H$; we explored $13.5 \le \mu \le 19.0$, which encompasses our closest bulge stars (assuming $A_H=0$) out to distance/extinction combinations that would remove a population from APOGEE's sample entirely. For each of the five APOGEE selection bins represented in our data set\footnote{$(J-K_s)_0 \ge 0.5$ and one of: $7.0 \le H < 11.0$, $7 \le H < 12.2$, $11 \le H < 12.2$, $11 \le H < 12.8$, or $12.2 \le H < 12.8$ \citep{zasowski2013,Zasowski_2017_apogee2targeting}}, we then counted the fraction of the $\log{g} \le 2.2$ RGB that fell within the bin.  
An example of these fractions is shown in Figure~\ref{fig:bias_fit}a, where the shading indicates $P_{\rm obs}$ as a function of [M/H] and $\mu$ for one APOGEE selection bin. A two-dimensional linear spline interpolation is used to store the pattern for each selection bin (Figure~\ref{fig:bias_fit}b).
Given the bin in which each real APOGEE star was selected, along with its metallicity, distance, and extinction (Sect.~\ref{subsec:dists_orbits}), we can account for the probability of finding it in our sample, relative to other stars.

For the second selection effect, we again use SSPs to compute, as a function of metallicity, the fraction of the RGB that lies beyond the DR16 ASPCAP synthetic spectra grid edge at $T_{\rm eff}=3000$~K (Figure~\ref{fig:limits}c). This fraction turns out to be well-represented by a simple function of metallicity (Figure~\ref{fig:bias_fit}c): 
\begin{equation}
P_{\rm [M/H]} = -0.04 \times \exp{\left(\frac{\rm [M/H]-0.46}{0.18}\right)} + 1,
\end{equation} 
where $P_{\rm [M/H]}=1$ means that the entire upper RGB (with $\log{g} \le 2.2$) lies within ASPCAP's range.

The final sample weight ($W_{\rm tot}$) for each star is taken as the inverse of the product of its observational and analysis probability, i.e., $W_{\rm tot} = 1/(P_{\rm obs} \times P_{\rm [M/H]})$. $P_{\rm [M/H]}$ is always large enough that it is not dominated by the [M/H] uncertainties. However, at small $P_{\rm obs}$, its value (and thus the value of $W_{\rm tot}$) is dominated by the uncertainty in $\mu$, which is roughly constant at $\sim$0.55, assuming 25\% distance uncertainties and 0.1~mag extinction uncertainties. Pairs of stars with $\Delta\mu \approx 0.55$~mag have differences in $P_{\rm obs}$ clustered around 0.1, suggesting $P_{\rm obs} \ge 0.1$ as a reasonable limit to ensure robust weighting.  Stars below this value ($\sim$8\% of the total bulge sample) are concentrated in the midplane, generally with $|Z| < 80$~pc; they cover the full $R_{\rm GC}$ span of our sample, but are much more heavily reddened than other stars at the same distance and latitude.  Their metallicity distribution is identical to the other stars (with $P_{\rm obs} > 0.1$) in the same latitude range, so our analysis and conclusions do not change at all if these stars with low probabilities are capped at $P_{\rm obs} = 0.1$, or indeed if they are excluded altogether. For simplicity, we set $P_{\rm obs} = 0.1$.

\begin{figure}
 \begin{center}
 \includegraphics[angle=0,trim=0in 0in 0in 0in, clip, width=0.45\textwidth]{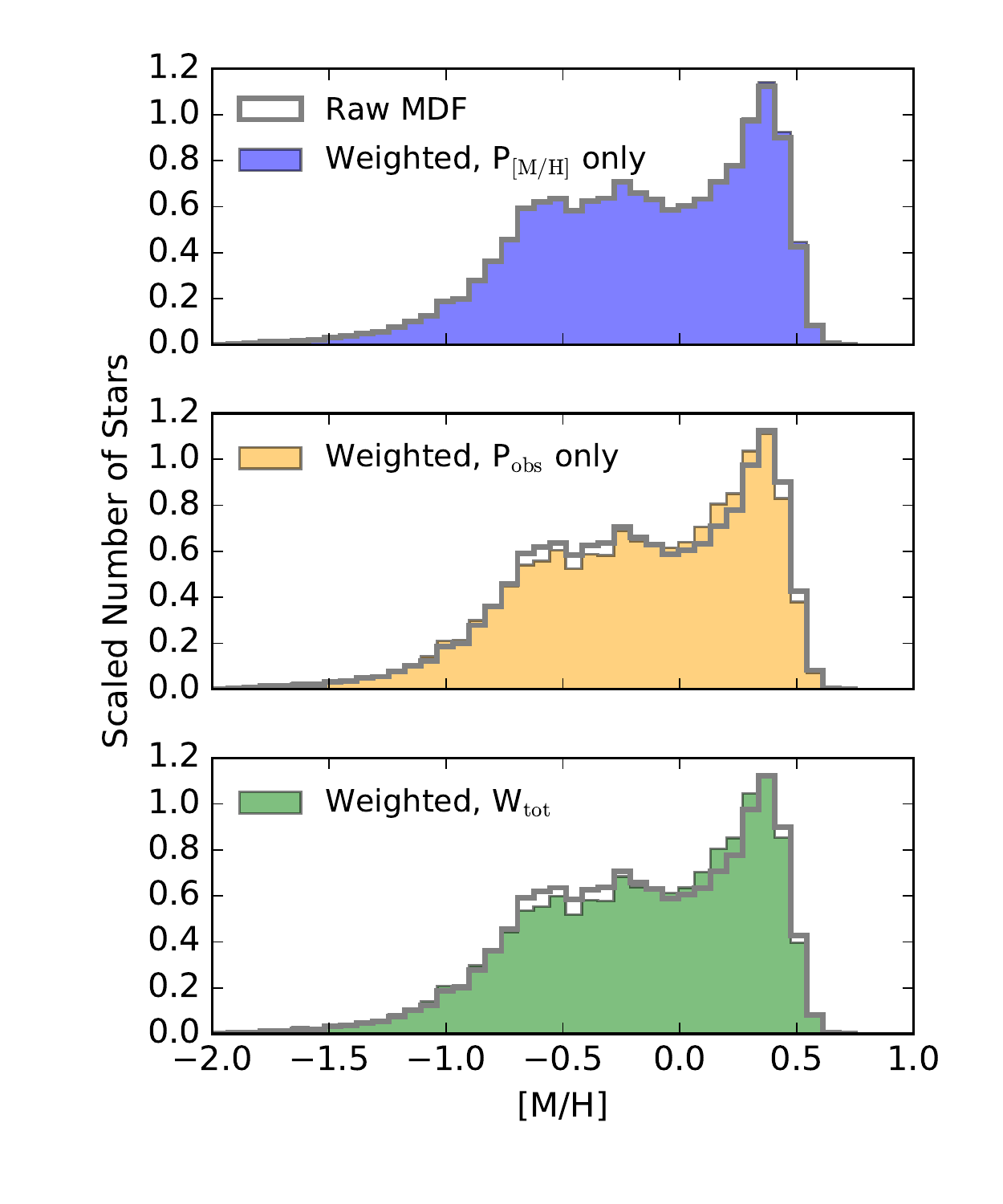} 
 \caption{ Selection- and metallicity-corrected MDFs for the total bulge sample (Sect.~\ref{sec:bias_corrections}), compared to the raw sample counts in gray. {\it Top panel:} MDF weighted by $P_{\rm [M/H]}$ alone. {\it Middle panel:} MDF weighted by $P_{\rm obs}$ alone. {\it Bottom panel:} MDF weighted by both $P_{\rm [M/H]}$ and $P_{\rm obs}$. In all cases, the weighting has a minimal impact on the global shape of the MDF.}
 \label{fig:mdf_corrected}
 \end{center}
 \end{figure}

The impact of applying these two types of weights is summarized in the MDFs shown in Figure~\ref{fig:mdf_corrected}. In short, the change in the MDF is minimal, suggesting that our bulge sample is not significantly affected by metallicity biases due to the survey selection function or to the effects of the current ASPCAP model grids.

\subsection{Exploring the bulge $R_{GC}$ limit}
\label{subsec:35kpc_justification}

Seen from the Sun, the Milky Way bulge has a boxy/peanut appearance, as initially established from the NIR light distribution and 2MASS star counts \citep[e.g.,][]{Dwek95, Lopez-Corredoira05}. Following early evidence from gas dynamics that indicates the presence of a non-axisymmetric rotating bar potential \citep{deVaucouleurs64,Rodriguez-Fernandez08}, studies using luminous stellar tracers (especially RC stars from massive photometric surveys) have characterized the boxy bar as a triaxial structure of length~$\sim$3.5~kpc, with axis ratios of 1:0.4:0.3 \citep{Rattenbury07,Robin12,wegg13}. Its near side points towards positive Galactic longitudes, with a position angle of $\sim25^\circ$ with respect to the Sun-Galactic center direction, and the asymmetric boxy isophotes are attributed to the effects of the near-end-on projection of a dynamically coherent bar.

Based on these considerations, many recent studies of bulge stellar populations have adopted a Galactocentric distance limit of $R_{\rm GC}=3.5-4$~kpc to define their bulge samples \citep{Ness13,Rojas-Arriagada17,Schultheis17}. Our aim in this section is to use the kinematical properties of our clean inner Galaxy sample (Sect.~\ref{subsec:sel_gen_clean_sample}) to assess the appropriateness of these limits and choose the best to use for this paper.

\begin{figure}
\begin{center}
\includegraphics[width=0.49\textwidth]{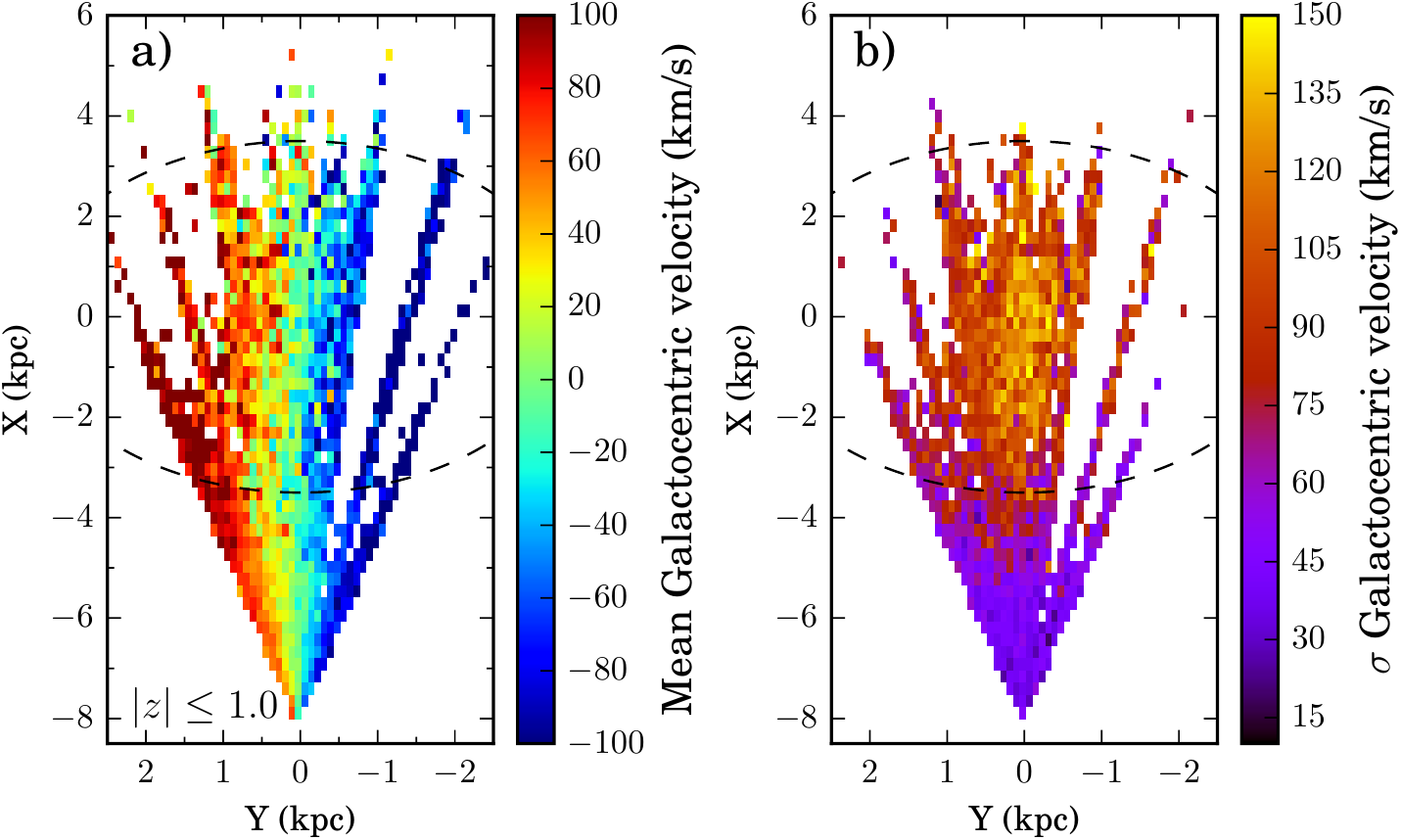}
\caption{Galactocentric velocity (${\rm V_{GC}}$) maps of the clean inner MW sample in the XY plane. They are computed over a $60\times60$ bins grid for stars with $|Z|\leq1.0$~kpc. \textit{Left panel:} mean Galactocentric velocity. Only pixels containing at least 4 samples are displayed. \textit{Right panel:} dispersion of the Galactocentric velocity. Only pixels containing at least 6 samples are displayed. The dashed black ellipses indicate in each panel the region defined by $R_{GC}=3.5$~kpc.}
\label{fig:35_just_kin}
\end{center}
\end{figure}

\begin{figure}
\begin{center}
\includegraphics[width=0.48\textwidth]{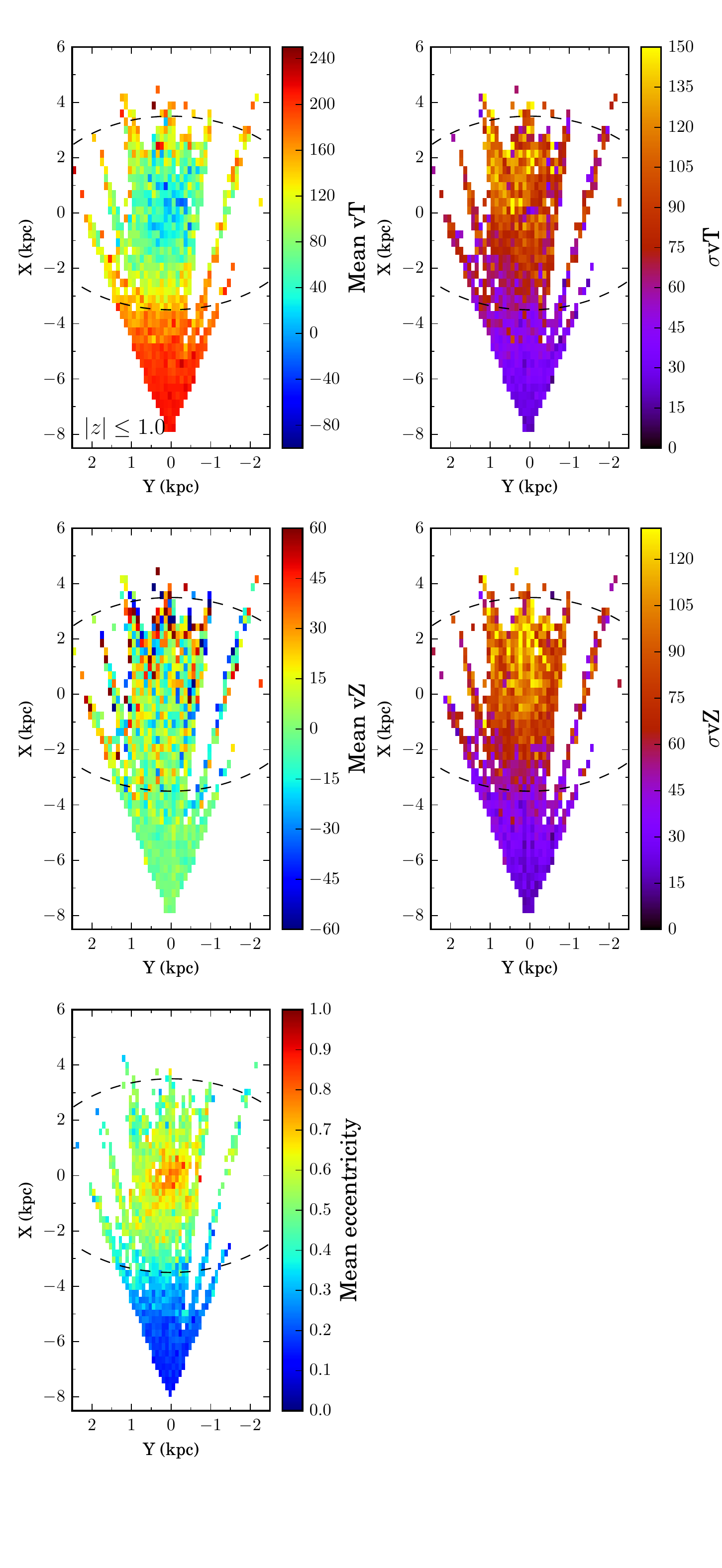}
\caption{The binned statistics of tangential velocity ($v_T$, upper row), vertical velocity ($v_z$, middle row) and eccentricity ($e$, lower row) are computed in the (X,Y) Galactic Cartesian plane. In each case, a $50\times50$ bins grid has been adopted, except for eccentricity in which case we adopt $60\times60$ bins. Only stars with $|Z|\leq1.0$~kpc are considered. In each panel, the red dashed-line ellipse depicts the $R_{GC}=3.5$~kpc limit.}
\label{fig:35_just_kin_dyn}
\end{center}
\end{figure}

In Figures~\ref{fig:35_just_kin} and \ref{fig:35_just_kin_dyn}, we show mean kinematical properties of our clean inner MW sample in the Galactocentric Cartesian ($X,Y$) plane. The Galactic Center is at $(X,Y)=(0,0)$~kpc, the Sun is at $(X,Y)=(-8.2,0)$~kpc, and the close end of the bar lies where $X<0$ and $Y>0$. Each property is shown for stars limited to $|Z| \le 1.0$~kpc. The dashed-line ellipses indicate $R_{\rm GC}=3.5$~kpc.

Figure~\ref{fig:35_just_kin} shows the Galactocentric radial velocity $V_{\rm GC}$ of the clean inner MW sample, with mean $\overline{V_{\rm GC}}$ in the left panel and dispersion $\sigma_{\rm VGC}$ in the right panel.
In the $\overline{V_{\rm GC}}$ panels, we see an overall uniform rotation pattern for stars at nearly all $R_{\rm GC}$, especially $R_{\rm GC} > 3.5$~kpc, where the pattern is particularly smooth. In contrast, the velocity dispersion ($\sigma_{\rm VGC}$) shows marked spatial patterns, with low dispersion (indicating coherent rotation) at $R_{\rm GC} \gtrsim 4$~kpc and higher dispersion (by a factor of two or more) at smaller radii, indicating more isotropic kinematics.

Figure~\ref{fig:35_just_kin_dyn} shows additional kinematic and orbital (eccentricity) properties for stars from the clean inner MW sample with good proper motions and orbit measurements (Sect.~\ref{subsec:dists_orbits}). These are defined as stars with uncertainties in both $\mu_\alpha\cos(\delta)$ and $\mu_\delta$ smaller than 0.5~mas~yr$^{-1}$ and uncertainties in tangential velocity ($v_T$, projected onto the plane), vertical velocity ($v_Z$), and eccentricity smaller than 40~km~s$^{-1}$, 20~km~s$^{-1}$, and 0.14, respectively. The error distributions of these quantities are all sharply peaked, so these limits only remove stars in the long high-error tails (beyond the 95th percentile of the distribution). In addition, as recommended by the {\it Gaia} consortium, we select stars with ${\rm RUWE<1.4}$, to ensure their astrometric solutions are reliable.

The top row of Figure~\ref{fig:35_just_kin_dyn} contains the $(X,Y)$ distribution of $\overline{v_T}$ (left) and $\sigma(v_T)$ (right), the middle row contains similar maps for $\overline{v_Z}$ and $\sigma(v_Z)$, and the bottom row contains a map for mean eccentricity $\overline{e}$. Together, these distributions tell a complementary story of a kinematical transition between the disc- and bulge-dominated regions. The disc is dominated by stars that are in roughly circular, coplanar rotation, with $v_T \approx 150-250$~km~s$^{-1}$, low dispersion in both $v_T$ and $v_Z$, and roughly uniform $e \lesssim 0.25$. Inside $R_{\rm GC} \sim 3.5$~kpc, however, rather sharply the kinematics become significantly less coherent, with $\sigma(v_T)$ and $\sigma(v_Z)$ greater than 100~km~s$^{-1}$ and mean eccentricity exceeding $\overline{e} = 0.6$. 

These trends indicate a relative dearth of circular, coplanar orbits and higher orbital isotropy in the inner Galaxy, and a transition between domination by the rotation-supported disc and the bar+pressure-supported bulge in a narrow region at $R_{\rm GC} \approx 3.5$~kpc. Based on this behavior, we adopt this as our spatial limit to select a sample of likeliest bulge stars (Sect.~\ref{subsec:sel_likely_bulge}).

\subsection{Selection of a bulge sample}
\label{subsec:sel_likely_bulge}

We use the results of Sect.~\ref{sec:bias_corrections}--\ref{subsec:35kpc_justification} to define our sample of bulge stars. Starting with the clean inner MW sample defined in Sect.~\ref{subsec:sel_gen_clean_sample}, we remove stars with $\log{g} > 2.2$~dex, to ensure our sample is adequately free of the selection and analysis effects of Sect.~\ref{sec:bias_corrections}. Then, we select stars with $R_{\rm GC}\leq3.5$~kpc as those in the kinematically-distinguished bulge/bar region explored in Sect.~\ref{subsec:35kpc_justification}. Finally, we focus for the rest of the analysis on the spatial variation of the MDF within the region bounded by $|\ell|\leq11^\circ$ and $|b|\leq13^\circ$, so we keep only stars inside those limits. The final bulge sample comprises 13,031 stars.

\begin{figure}
\centering
\includegraphics[width=0.48\textwidth]{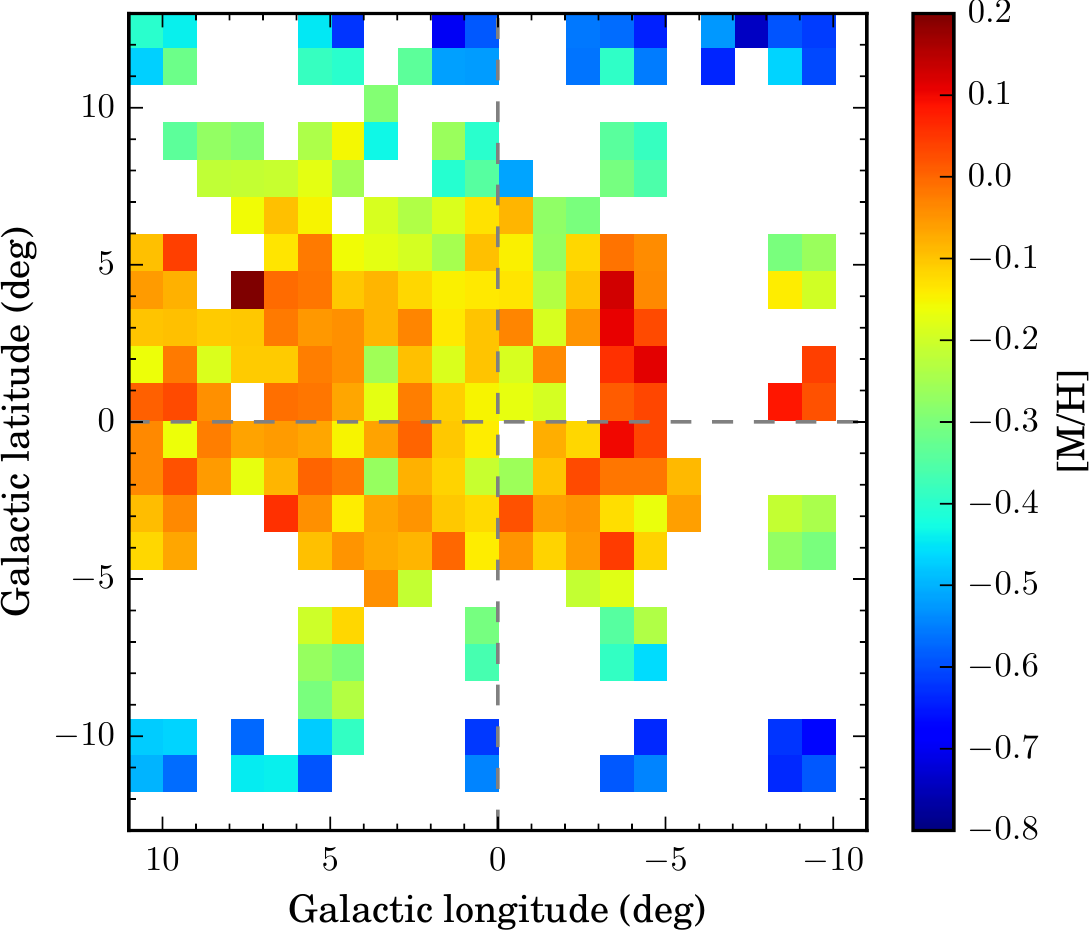}
\caption{Mean metallicity map of our bulge sample. The mean metallicity is computed over a $22\times22$ grid within $|\ell|\leq11^\circ$ and $|b|\leq13^\circ$. Only pixels with more than 7 stars are shown.}
\label{fig:mean_metallicity_map}
\end{figure}

Figure~\ref{fig:mean_metallicity_map} shows the mean metallicity map (in $\ell,b$) of the bulge sampled by the APOGEE bulge stars. The mean metallicity was computed over a $22\times22$ grid in the $|\ell|\leq11^\circ$, $|b|\leq13^\circ$ region, displaying only pixels containing more than seven stars. The dominant gradient is in latitude, with average metallicity increasing towards the midplane. In addition, although the sampling of APOGEE data is less dense beyond 5$^\circ$ from the Galactic plane, the general pattern appears symmetric around $b=0^\circ$. Beyond $|b|>10^\circ$, however, a weak longitudinal gradient is visible in the map, with the mean population becoming more metal-rich towards positive longitudes. As we shall see, this region is dominated by stellar populations other than bulge stars, and it is at the border of the region we are sampling. Consequently, we do not quantify the significance of this effect. Qualitatively, this mean metallicity map compares quite well with the photometric map of \citet{Gonzalez13} derived from VISTA Variables in the V\'ia L\'actea survey \citep[VVV, ][]{Minniti10} data, and shows the vertical metallicity gradient (see also Fig.~\ref{fig:gmm_strip_summary}).

\section{The MDF}
\label{sec:mdf}

\begin{figure}
\centering
\includegraphics[width=0.48\textwidth]{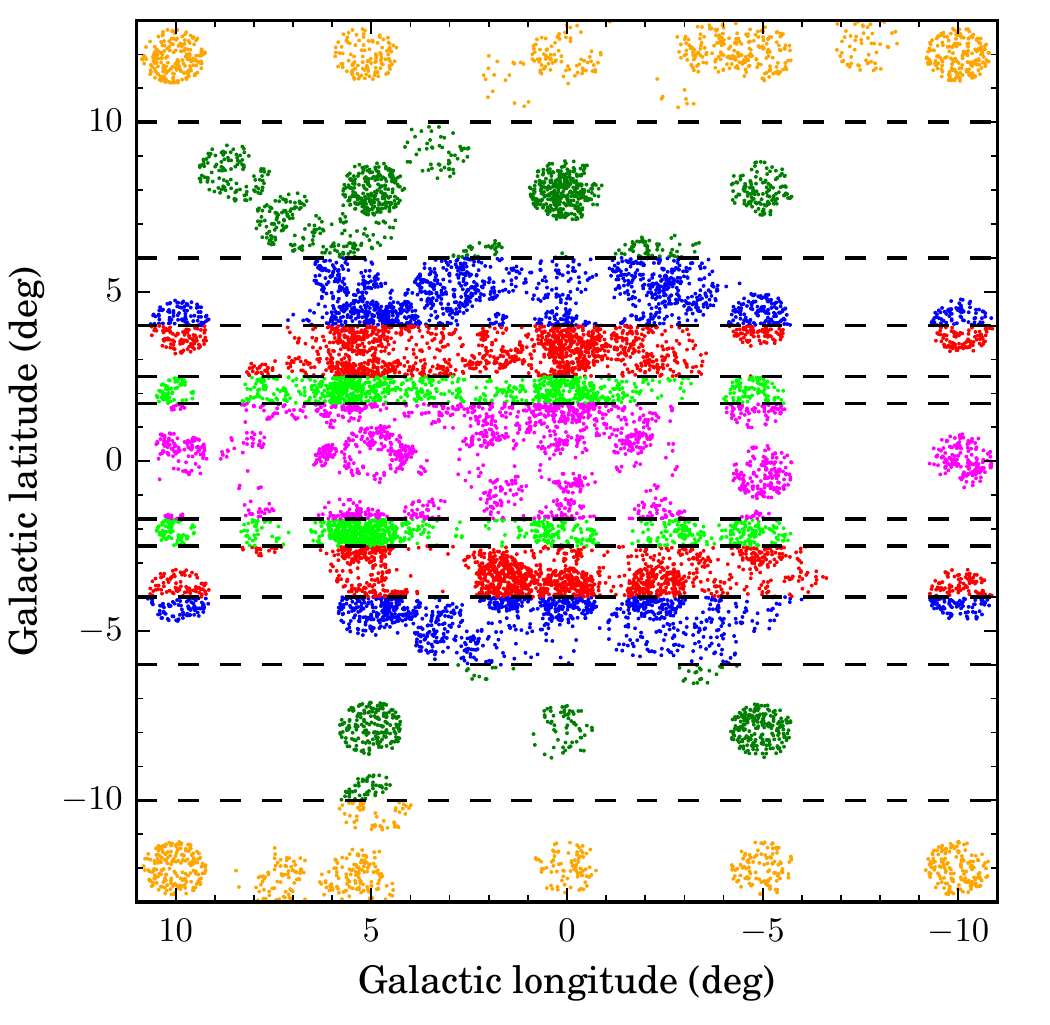}
\caption{Distribution of our bulge sample of 13,031 stars in $(l,b)$. The horizontal black dashed lines define our latitude strips, with a binning pattern that is symmetric about the midplane (Section~\ref{sec:mdf}). Points are color-coded by latitude strip.}
\label{fig:lb_strips}
\end{figure}

Figure~\ref{fig:lb_strips} shows the $(l,b)$ distribution of our bulge sample, emphasizing (along with Figure~\ref{fig:sample}) that APOGEE  preferentially samples the region within 5$^\circ$ of the Galactic plane, even with stars at $R_{\rm GC} \leq 3.5$~kpc. This region has been largely avoided by large-scale optical spectroscopic surveys because of the severe limitations imposed by the high dust extinction. Thanks to its NIR $H$-band sensitivity, APOGEE is less affected by this extinction and offers the opportunity to systematically explore the nature of the MDF with good statistics across dozens of square degrees in the innermost parts of the MW.

Numerous bulge studies have shown that the proportion of metal-rich to metal-poor stars decreases as a function of the distance ($|b|$ or $|Z|$) from the midplane \citep{zoccali08,Ness13,Rojas-Arriagada17}. 
As we also focus here on the latitude dependence of the MDF, we split our bulge sample into latitude strips that extend across the full longitude range and are symmetric around $b=0^\circ$ (Figure~\ref{fig:lb_strips}); we combine stars into strips of $|b|$ after confirming the MDFs of the matched strips at $\pm b$ are statistically identical (see also the symmetry of Figure~\ref{fig:mean_metallicity_map}).
The varying widths of the strips are chosen to maximize the resolution in $|b|$ while still ensuring more than 1600 stars per strip, each of which is shown in a different color in Figure~\ref{fig:lb_strips}. The three outermost strips include stars spanning areas previously studied by optical spectroscopic surveys (e.g., ARGOS, GIBS, and GES at $b<-4^\circ$).

The goal of this paper is to study in greater detail the {\it shape} --- especially the ``peakiness'' --- of the MDF and how that shape changes with latitude in the relatively poorly explored inner few degrees of the bulge, all enabled by the large size and wide angular distribution of the APOGEE sample. In Section~\ref{sec:mdf_properties}, we describe the qualitative properties of the MDFs in our six bins of $|b|$.  In Sect.~\ref{sec:mdf_gmm}--\ref{subsec:kin_mdf_comps}, we explore statistical decompositions of these MDFs, the stability of these decompositions with $|b|$, and the kinematical properties of the resulting components.  In Sect.~\ref{sec:nmf}, we use an alternate decomposition method to highlight how different methods may affect our interpretation of bulge ``components''.

\subsection{MDF Properties}
\label{sec:mdf_properties}

\begin{figure}
\centering
\includegraphics[width=0.48\textwidth]{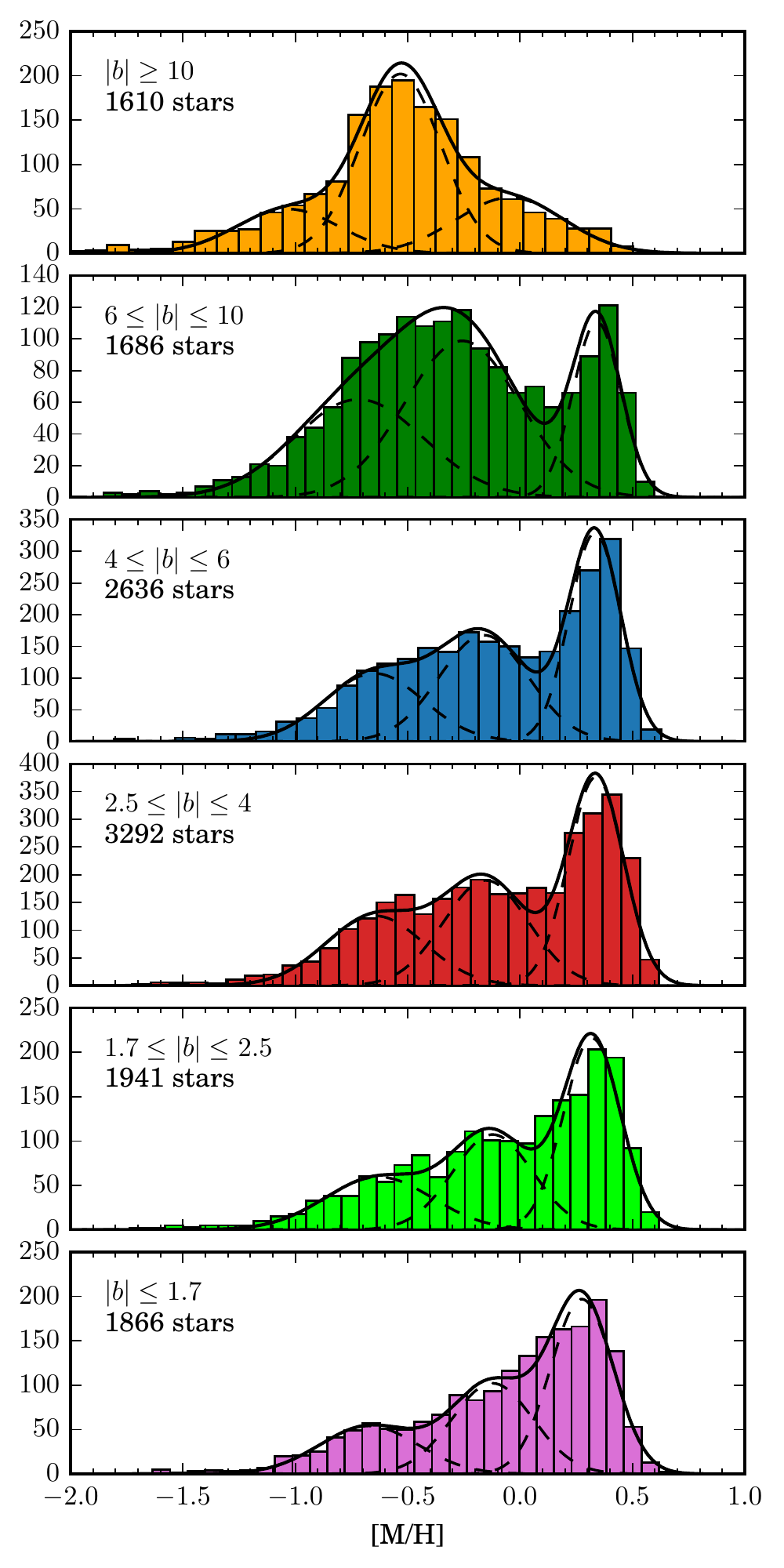}
\caption{Metallicity distribution functions of the latitude strips shown in Figure~\ref{fig:lb_strips}, with the same color scheme. The black dashed and solid lines indicate the individual and summed components, respectively, of the optimal GMM modeling of the data's density distribution (Section~\ref{sec:mdf_gmm}).}
\label{fig:mdf_strips}
\end{figure}

Figure~\ref{fig:mdf_strips} shows the MDFs of the spatial bins defined above, along with Gaussian mixture components that are discussed in Sect.~\ref{sec:mdf_gmm}.

The MDF of the outermost spatial (folded) bin at $|b|\geq10^\circ$ is dominated by a single peak at $\rm [M/H] \sim -0.5$~dex (top panel of Figure~\ref{fig:mdf_strips}). Its overall shape is highly symmetric, but not Gaussian, with pronounced tails extending towards the metal-rich and metal-poor ends. The location of the dominant peak is too metal-rich to be attributed to halo stars (although halo stars might be present in the metal-poor tail of the distribution), but it is consistent with the presence of thick disc stars dominating the line-of-sight mix of stellar populations at high Galactic latitude (for reference, stars located at $|b|=10^\circ$ are at $|Z|\sim1.45$~kpc from the midplane at a $d=8.2$~kpc distance to the bulge).

At $6^\circ \leq |b| \leq 10^\circ$ the MDF appears bimodal, dominated by a broad metal-poor distribution with a peak at $\rm [M/H] \sim -0.4$~dex. The second prominent peak is narrower and more metal-rich, centered at $~+0.4$~dex. A similar, but less peaky, bimodal distribution is seen in the stars at $4^\circ\leq|b|\leq6^\circ$. In this case, the metal-rich peak is dominant and well defined, while metal-poor stars are present in a flatter distribution. The spatial region covered by these two strips has been previously sampled by optical spectroscopic surveys such as GIBS and GES (at negative latitudes). The overall picture proposed from these surveys is that the bulge MDF is intrinsically bimodal, with the relative proportion of metal-rich stars increasing as closer to the midplane. The visual inspection of the MDFs from our data over the same spatial region seems to qualitatively confirm this picture.

Inside $|b|<4^\circ$, the MDF becomes increasingly dominated by the metal-rich peak (lower three panels of Figure~\ref{fig:mdf_strips}). As $|b|$ gets smaller, the metal-poor distribution becomes increasingly flat, resembling more of a heavy tail to the metal-rich peak; upon close examination, this tail appears to comprise two wide peaks, separating around $\rm [M/H] \sim -0.4$~dex. This separation is weaker in the innermost strip ($|b| \leq 1.7^\circ$).

\begin{figure}
\centering
\includegraphics[width=0.48\textwidth]{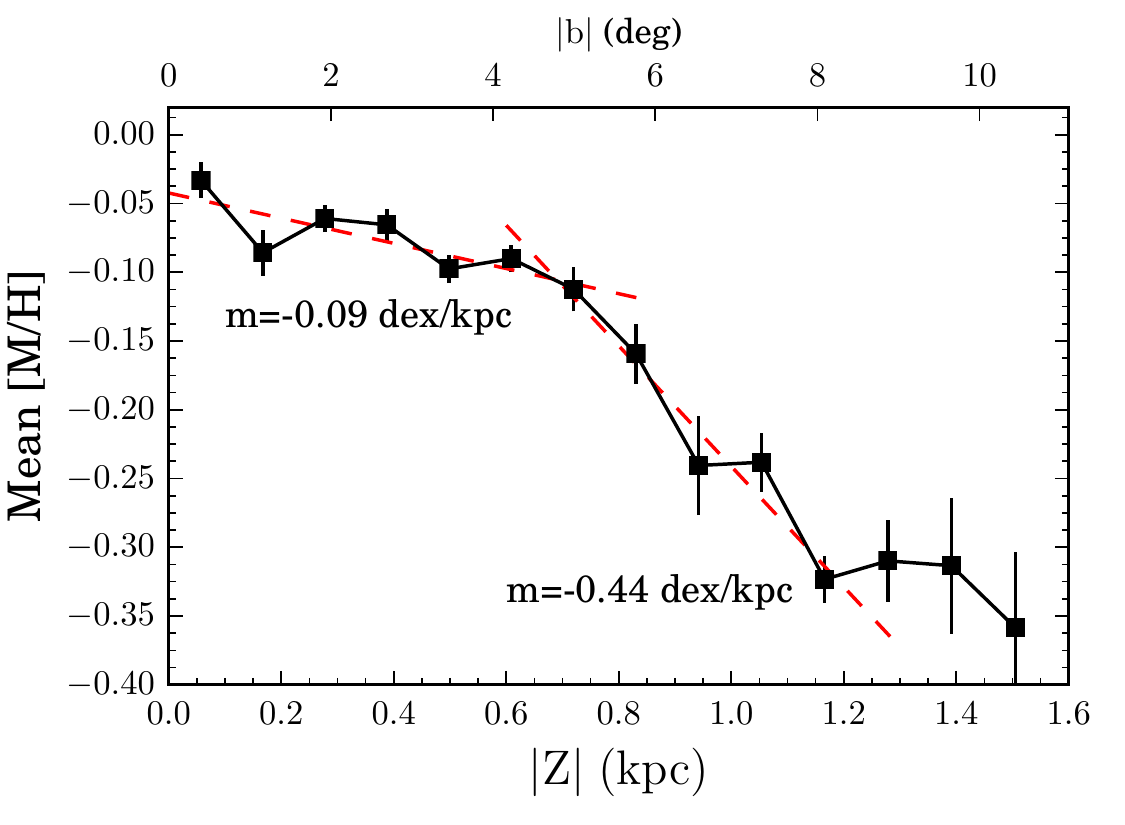}
\caption{Vertical metallicity gradient computed from the mean metallicity of stars in a number of narrow latitude intervals. Angular distances (in the additional top axis) has been converted to Z by assuming a projection on a a plane at 8~kpc (to be consistent with other studies and allow comparison). The gradient has been estimated in two independent regions, as indicated by the red dashed lines.}
\label{fig:vertical_gradient}
\end{figure}

In Fig.~\ref{fig:vertical_gradient} we divide the bulge sample in several narrow latitude strips in order to compute the vertical mean metallicity gradient. The angular distances were converted to spatial vertical distances by assuming all stars in each bin projected in a plane at 8~kpc. This is done in order to be consistent with other studies and allow comparison. The vertical gradient  has been measured in previous studies as a summary quantity of the vertical variation of the composition of lines of sight towards the bulge. The resulting gradient appears to have two slopes: a flatter one of $-0.09$~dex/kpc within 0.7~kpc ($|b|\sim5^\circ$) from the midplane, and a steeper one of $-0.44$~dex/kpc from there out to 1.2~kpc ($|b|\sim8^\circ$). Beyond that point, the slope becomes flat but noisy. The slope we measure beyond $|b|\sim5^\circ$ is lower than the value reported over a similar region ($-12^\circ \leq b\leq-4^\circ$) by \citet{zoccali08} and \citet{Rojas-Arriagada17} ($-0.24$~dex/kpc, both). The shallower value reported in these works may be driven by the inclusion of outer fields, which as we see in our data, are flatter but noisy. On the other hand, the inner flattening of the metallicity gradient is in agreement with previous suggestions from the analysis of smaller samples of high resolution NIR spectra \citep{Rich07,Rich12,Schultheis19}.

This spatial variation of the shape of the MDF reflects a complex mix of stellar populations present in the bulge. The identification and characterization of this complexity has been a key outcome from the optical surveys exploring the southern bulge region at $b \lesssim -4^\circ$. The presence of at least two separate ``populations'' (one metal-rich and one metal-poor) with different spatial distributions, different kinematics, and possibly different origins is argued to drive the variable behavior of stellar distributions, including the MDF.

\subsection{MDF Decomposition with Gaussian mixture modeling (GMM)}
\label{sec:mdf_gmm}

\begin{figure}
\centering
\includegraphics[width=0.48\textwidth]{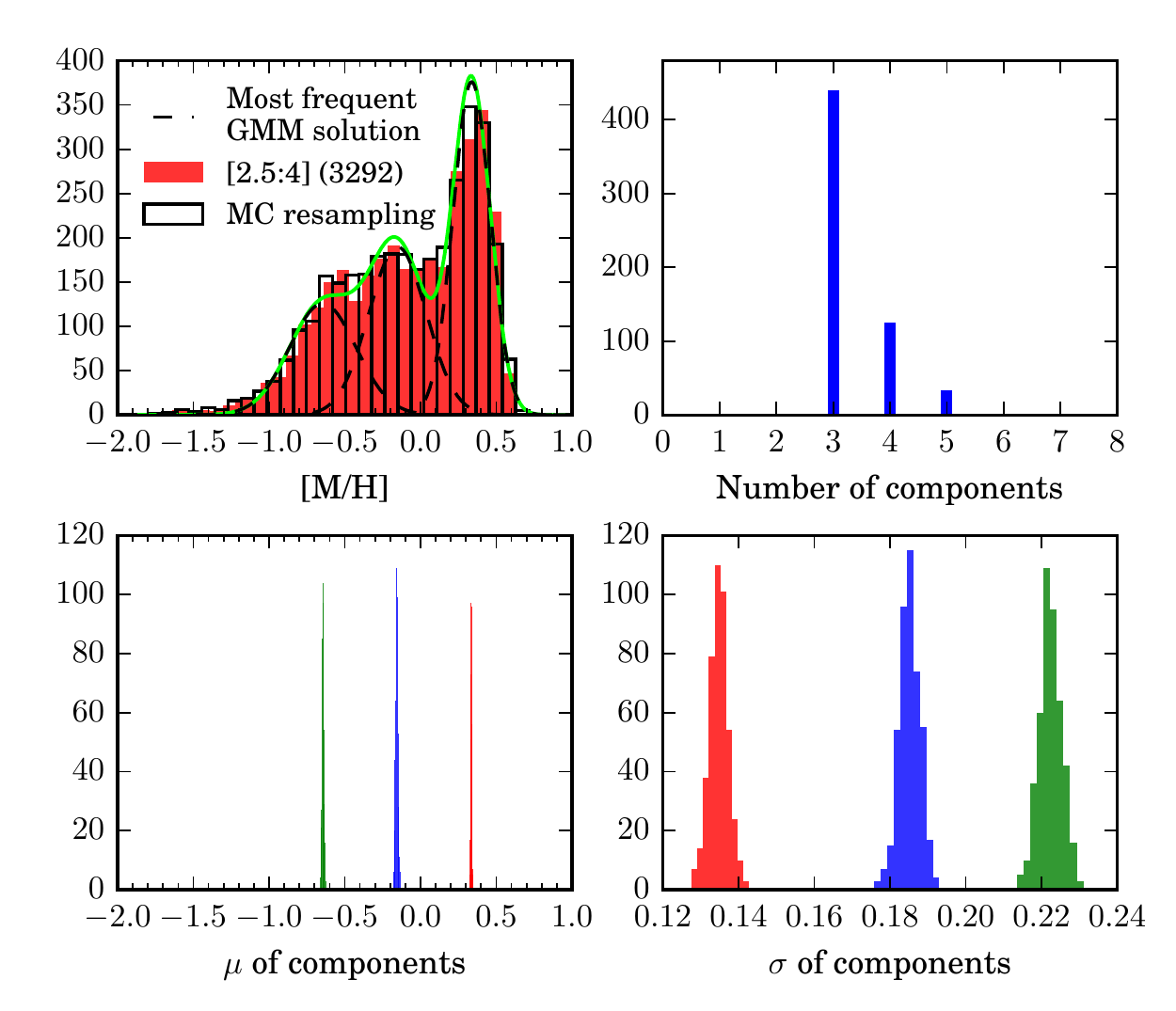}
\caption{Example of the Monte Carlo analysis implemented to assess the stability of the GMM solution of the MDF. \textit{Top left panel:} the red histogram stands for the observed MDF in a given latitude strip indicated in the legend, as well as the number of stars in that strip. The black step histogram depicts one random sampling of the observed MDF given metallicity measurement uncertainties. The black dashed lines show the individual components, while the solid green line total profile of the optimal trimodal Gaussian mixture. \textit{Top right panel:} frequency distribution of the optimal number of Gaussian components found from the GMM runs on the 600 Monte Carlo resamplings of the observed MDF. \textit{Bottom panels:} frequency distributions of the centroids (left) and widths (right) of each Gaussian component in the preferred trimodal parameterization.} 
\label{fig:gmm_mc_example}
\end{figure}

Identifying and separating by chemistry these ``populations'' described above has largely been done by assuming the intrinsic shapes of their underlying distributions can be described by simple functions (most commonly, Gaussians) that do not have a large overlap.  Expanding on this tradition, we use Gaussian Mixture Modeling (GMM) to decompose the $|b|$-dependent MDFs and study not only the structure of the distribution, but also how robust this structure is to uncertainties in the data, how robust it is to the stochasticity of GMM decomposition, and how the peakiness itself depends on $|b|$.

To this end, we adopted a Monte Carlo resampling approach, which is illustrated in Fig.~\ref{fig:gmm_mc_example}. Given an observed MDF (red histogram, top left panel), we draw 600 resamplings of the individual stellar metallicities by assuming a Gaussian variability of 0.05~dex \citep[approximately APOGEE's nominal metallicity measurement uncertainty; e.g., ][]{Holtzman_2018_dr13dr14apogee, Jonsson_2018_dr13dr14abundances}. An example of the MDF resulting from one such resampling is shown as a black empty histogram in the top left panel of Fig.~\ref{fig:gmm_mc_example}. 

A GMM parameterization was computed for each of the 600 resamplings, considering mixtures with $N=1-6$ components, and using the Bayesian Information Criterion (BIC) to identify the optimal model. The frequency of the optimal number of components over the whole set of resamplings was examined to see if a preferred model complexity emerged (top right panel). A trimodal solution is found to be the optimal one in the majority of the cases ($\sim75\%$ in this example). Thus, we choose the trimodal mixture as the stable optimal parameterization of this MDF density structure. 

The lower panels of Fig.~\ref{fig:gmm_mc_example} show the distributions of the centroids and widths of the individual Gaussian components in the trimodal solutions. The narrowness of these distributions indicate that the component parameters are robust over the set of Monte Carlo resamplings. The optimal trimodal Gaussian mixture is shown on top of the observed MDF in the top left panel. Dashed black lines denote the individual components, while the solid green line shows the total density distribution. This distribution appears to be a fairly good description of the density distribution of the observed data at this latitude.

\begin{table}
\begin{center}
\begin{tabular}{c|cccccc}
Galactic Latitude               &   N=1 & N=2 &  N=3 &  N=4 &  N=5 &  N=6 \\[1pt]
\hline
\hline
$10^\circ \leq |b| $         &  0.00 & 0.00 & 0.99 & 0.01 & 0.00 & 0.00 \\
$6^\circ \leq |b| < 10^\circ$    &  0.00 & 0.37 & 0.56 & 0.05 & 0.02 & 0.00 \\
$4^\circ \leq |b| < 6^\circ$     &  0.00 & 0.11 & 0.47 & 0.34 & 0.09 & 0.00 \\
$2.5^\circ \leq |b| < 4^\circ$   &  0.00 & 0.00 & 0.75 & 0.19 & 0.06 & 0.00 \\
$1.7^\circ \leq |b| < 2.5^\circ$ &  0.00 & 0.00 & 0.56 & 0.38 & 0.07 & 0.00 \\
$|b| < 1.7^\circ$ &  0.00 & 0.00 & 0.93 & 0.07 & 0.00 & 0.00 \\
\end{tabular}
\end{center}
\caption{Relative frequency of the preferred number of GMM components used to model the MDF over the 600 Monte Carlo resamplings in each latitude strip.}
\label{tab:freq_components_mc_mdf}
\end{table}

We applied the procedure described above to the observed MDFs in each of our latitude strips (Fig.~\ref{fig:mdf_strips}). 
The components of the optimal model are shown on top of each MDF in dashed lines, with their sum as the solid line.
Despite the changing shape of the MDF with latitude, {\it in all cases the preferred GMM solution is trimodal}.
This can be seen in Table~\ref{tab:freq_components_mc_mdf}, which contains the relative frequency of solutions with  $N=1-6$ components over the whole set of Monte Carlo resamplings of each latitude strip.
Table~\ref{tab:weight_coeffs} contains the parameters of the optimal Gaussian mixture for each of the latitude strips (Fig.~\ref{fig:mdf_strips}).

\begin{table*}
\centering
\begin{tabular}{l|ccc|ccc|ccc}

    \multirow{2}{*}{Strip} &
      \multicolumn{3}{c|}{metal-poor} &
      \multicolumn{3}{c|}{metal-intermediate} &
      \multicolumn{3}{c}{metal-rich} \\
    & $\mu$ & $\sigma$ & $w$ & $\mu$ & $\sigma$ & $w$ & $\mu$ & $\sigma$ & $w$ \\
\hline
\hline
$10\leq|b|$         & $-1.04$ &  0.23 & 0.18 & $-0.53$ &  0.18 & 0.58 & $-0.04$ &  0.24 & 0.24 \\
$6\leq|b|\leq10$    & $-0.72$ &  0.29 & 0.33 & $-0.26$ &  0.26 & 0.46 & $0.34$ &  0.11 & 0.22 \\
$4\leq|b|\leq6$     & $-0.65$ &  0.22 & 0.25 & $-0.16$ &  0.20 & 0.35 & $0.33$ &  0.11 & 0.40 \\
$2.5\leq|b|\leq4$   & $-0.64$ &  0.22 & 0.25 & $-0.15$ &  0.19 & 0.33 & $0.34$ &  0.12 & 0.42 \\
$1.7\leq|b|\leq2.5$ & $-0.63$ &  0.24 & 0.23 & $-0.12$ &  0.18 & 0.32 & $0.32$ &  0.13 & 0.45 \\
$|b|\leq1.7$        & $-0.68$ &  0.22 & 0.20 & $-0.13$ &  0.18 & 0.32 & $0.27$ &  0.14 & 0.48 \\
\\[1pt]
\end{tabular}
\caption{The mean [M/H] ($\mu$), width ($\sigma$) and relative weights ($w$) of the three individual components found to be the preferred and stable GMM solution in the MDF of our latitude strips.}
\label{tab:weight_coeffs}
\end{table*}

The metal-rich peak, which is visually apparent in all the MDFs, corresponds closely to the consistently-narrowest of the GMM components (except in the outermost latitude bin at $|b|\geq10$, which as mentioned before, is likely dominated by contamination). Interestingly, the broad metal-poor distribution identified in the qualitative assessment of the MDF shape (Sect.~\ref{sec:mdf_properties}) appears to be best described by two Gaussian components. This is the case even in the strips where visual inspection may suggest a single metal-poor peak. In the rest of this analysis, we will refer to these components as metal-rich, metal-intermediate, and metal-poor. As we shall argue, this nomenclature does not necessarily imply physically different structures/populations, but rather distinct {\it data} components needed to parameterize the density structure of the observed MDF.

\begin{figure}
\centering
\includegraphics[width=0.48\textwidth]{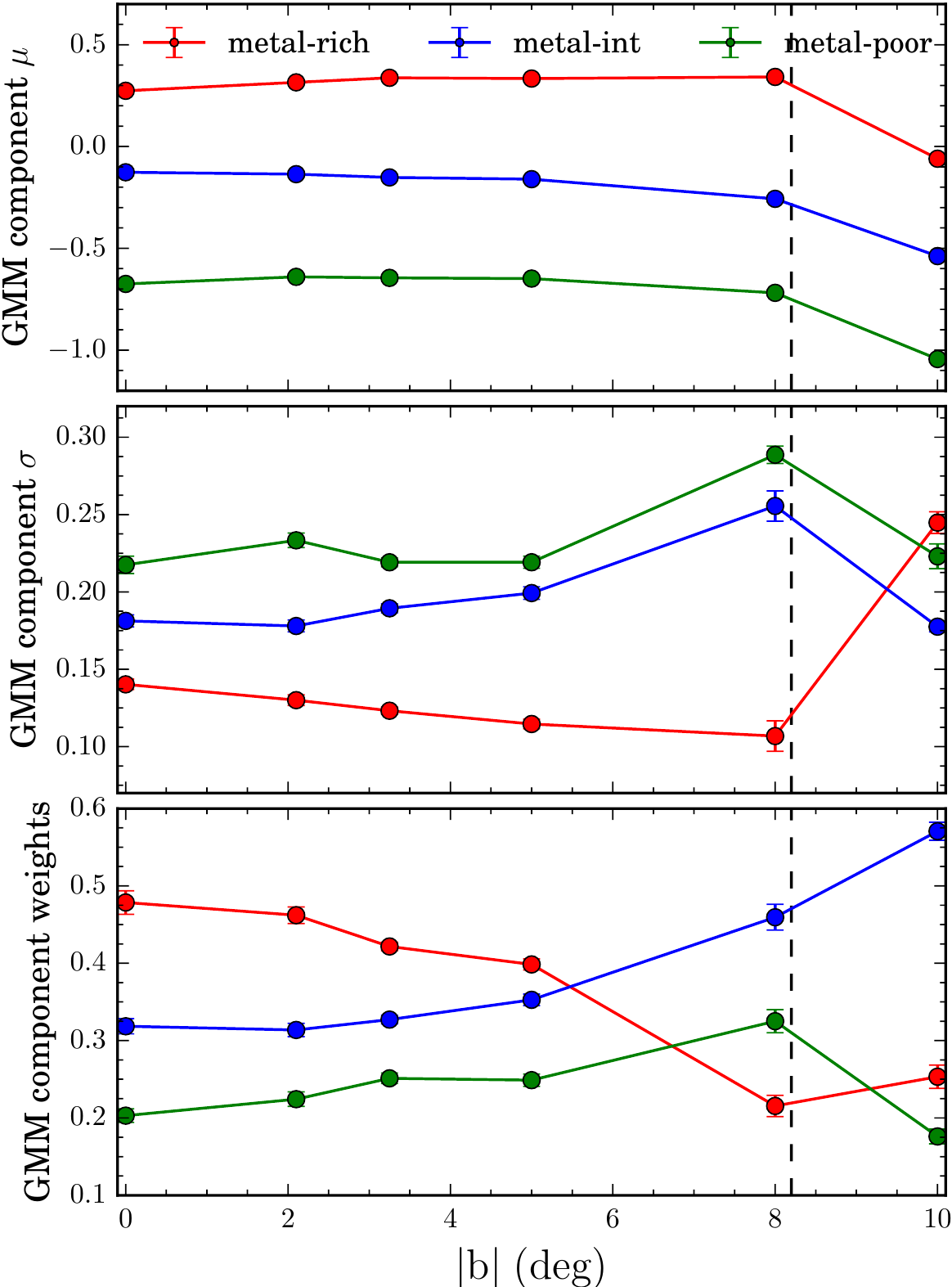}
\caption{Summary plots of the GMM analysis on the latitude-separated MDFs. The curves display the variation as a function of the distance from the midplane of the individual Gaussian parameters of the preferred trimodal mixture: centroids ($\mu$, top panel), widths ($\sigma$, middle panel), and relative weights ($w$, bottom panel). In all panels, the vertical black dashed line separates the region seemingly dominated by three similar bulge components (i.e. $|b| < 8^\circ$) from that dominated by different components, possibly describing the halo$+$thick disc (Sect.~\ref{sec:mdf_gmm}).}
\label{fig:gmm_strip_summary}
\end{figure}

The narrowness of the probability distribution functions for the centroid $\mu$ of each component (e.g., bottom left panel of Fig.~\ref{fig:gmm_mc_example}), and the constancy of $\mu$ with $|b|$ (e.g., top panel of Fig.~\ref{fig:gmm_strip_summary}), suggest that the metallicities of the components do not themselves depend on distance from the Galactic plane.
In this sense, the shape of the MDF is determined by the combination of components whose relative weights change systematically, with metal-rich stars becoming progressively important closer to the midplane. This observation is consistent with the qualitative picture drawn from optical surveys.

We note that although the MDF of the sample at $|b|\geq10^\circ$ is also best represented by a trimodal mixture, the individual components look different from those at other latitudes. The most prominent component peaks at ${\rm[M/H]}\sim-0.5$~dex, which is located between the metal-intermediate and metal-poor peaks of the other latitude samples. A similar offset is observed for the two other components of this mixture. This reinforces our previous hypothesis that our highest latitudes are dominated by stars from the thick disc and halo.

Fig.~\ref{fig:gmm_strip_summary} contains a summary of the individual component means ($\mu$), widths ($\sigma$), and relative weights as a function of angular distance from the midplane. The top panel highlights the constancy of the metallicity centroids of the individual components over the whole $0\leq|b|\lesssim8^\circ$ region, only deviating in the outermost bin at $|b|\geq10$. A similar trend is seen for the widths of the components (second panel from top), although with larger deviations than the centroids. 

The trend of the relative weight of components with $|b|$ (third panel) suggests that the global shape of the MDF is largely determined by the variation of the relative weights of the metal-rich and metal-intermediate components. In this sense, the metal-poor component accounts for a smaller (and relatively constant) proportion of the data density, with only a hint of a modest enhancement at higher latitudes. The inversion in the relative importance of the metal-rich and metal-intermediate components happens in the range $|b|\sim5^\circ-7^\circ$, with the ratio becoming stable at the innermost $2-3$ degrees.

\begin{figure*}
\centering
\includegraphics[width=0.90\textwidth]{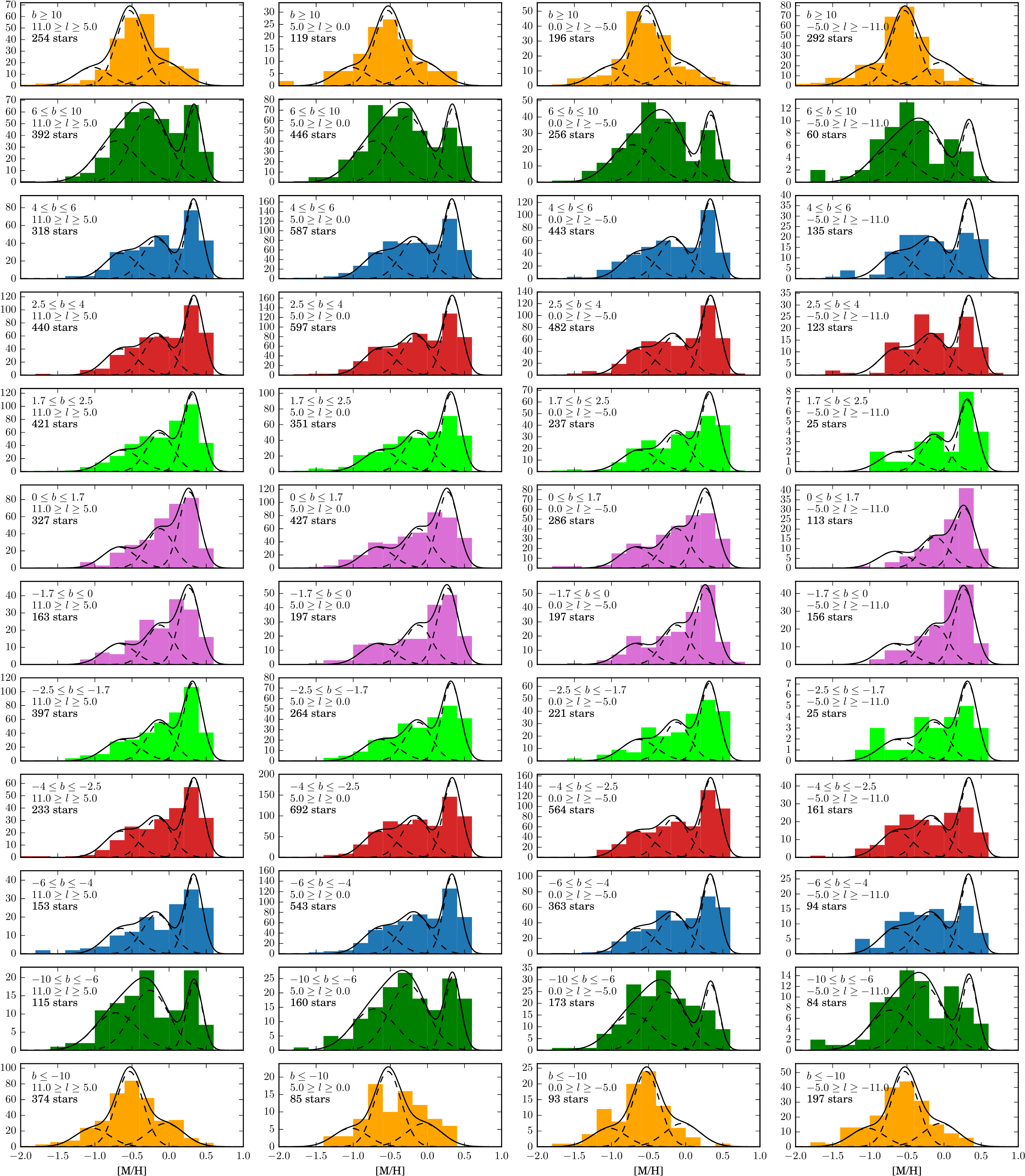}
\caption{Variations of the bulge MDF in binned regions in $(l,b)$. Stars in each latitude strip (see Fig.~\ref{fig:lb_strips}) are split into four longitudinal ranges: $5^\circ\le l\leq11^\circ$, $0^\circ\le l\leq5^\circ$, $-5^\circ\le l \leq0^\circ$, and $-11^\circ\le l \leq -5^\circ$. The MDFs of the resulting samples are distributed in the figure to preserve their relative position in the $(l,b)$ plane, as shown in Fig.~\ref{fig:lb_strips}. The Galactic longitude/latitude range and number of stars are quoted in each panel. The best-fit GMM to the full sample in each latitude strip is overplotted with dashed and solid black lines (individual components and full mixture, respectively). } 
\label{fig:mdf_global_small_bins}
\end{figure*}

In Fig.~\ref{fig:mdf_global_small_bins}, we compare the $|b|$-dependent best-fit GMM solutions to smaller subsamples separated around the midplane (i.e., using $b$ instead of $|b|$) and further divided by Galactic longitude: $-11^\circ\leq l\leq-5^\circ$, $-5^\circ\leq l\leq0^\circ$, $0^\circ\leq l\leq5^\circ$ and $5^\circ\leq l\leq11^\circ$. The arrangement of the MDFs of these angular pixels in Fig.~\ref{fig:mdf_global_small_bins} is consistent with the layout in Fig.~\ref{fig:lb_strips}. The distributions are noisier, as expected, but the number of stars per bin is still typically $>$100.
As a visual reference, in each panel, we show the best-fit GMM to the full sample in each latitude strip with dashed (individual components) and solid black (summed mixture) lines. Stars from low to super-solar metallicity are observed in all individual MDFs across the whole $(l,b)$ region.
At fixed $l$, the MDF of fields symmetric about the midplane (i.e., at $\pm b$) appear consistent with each other.

On the other hand, if one compares the MDF of individual fields along a latitude strip, one sees that the strength of the metal-rich peak finds better agreement at positive longitudes to the respective whole-latitude-bin GMM model. This asymmetry is clearer in the strips beyond $|b|=2.5^\circ$. The trend may be explained by the asymmetry introduced by the bar, whose major axis lies at an angle to the Sun-Galactic Center line and whose near end is located at positive longitudes (the sight-lines at negative longitudes must extend to larger distances to reach the metal-rich dominated bar). This is consistent with the association of metal-rich stars with the bar (also see \citeauthor{Wegg19} \citeyear{Wegg19} and Hasselquist et al. in prep). We emphasize that these asymmetries manifest themselves only as changes in the relative component weights of the observed (noisy) MDF with respect to the latitude best fit GMM, not the centers or widths of the components.

In summary, our GMM analysis shows that, when based on statistically significant samples, the density structure of the bulge MDF is optimally and robustly parameterized by a trimodal Gaussian mixture. The metallicity of the metal-rich, metal-intermediate and metal-poor components remains nearly constant with latitude at approximately $-0.66$, $-0.17$, and $+0.32$~dex, respectively. The strongest variation of the MDF shape is in the vertical direction, and appears mostly driven by the variation of the relative weight of the metal-rich and metal-intermediate components. The metal-poor component seems to account for a relatively constant and low fraction of the total MDF density, regardless of the distance to the midplane.

\subsection{Kinematic properties of the MDF GMM components}
\label{subsec:kin_mdf_comps}

\begin{figure*}
\centering
\includegraphics[width=0.96\textwidth]{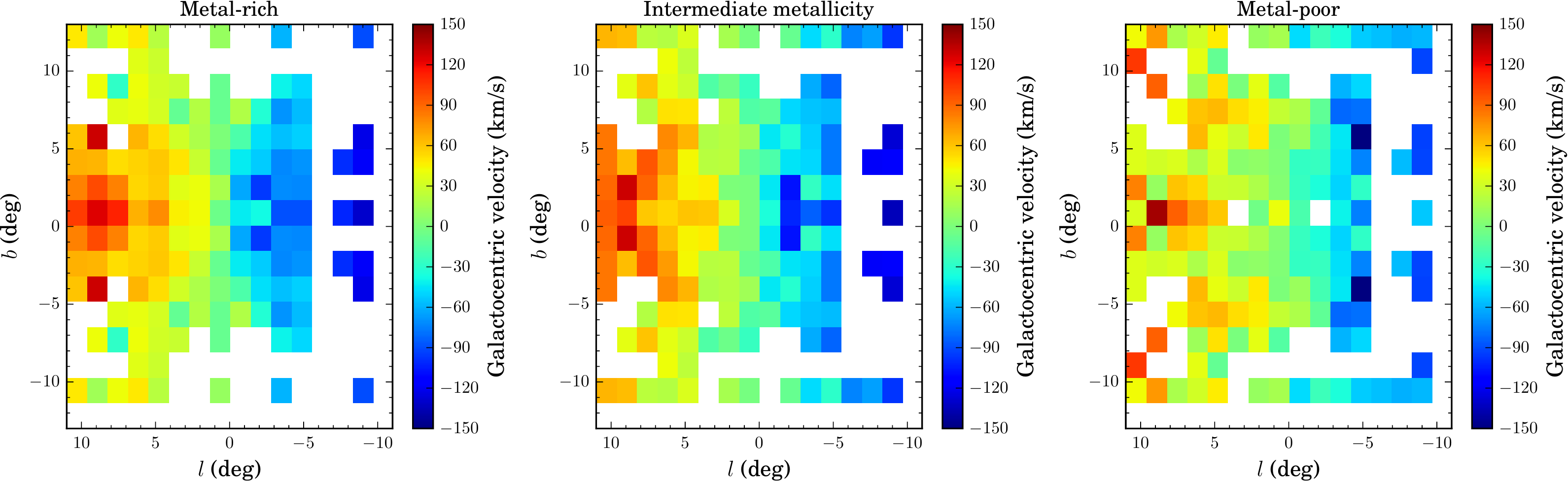}
\vspace{0.5cm}

\includegraphics[width=0.96\textwidth]{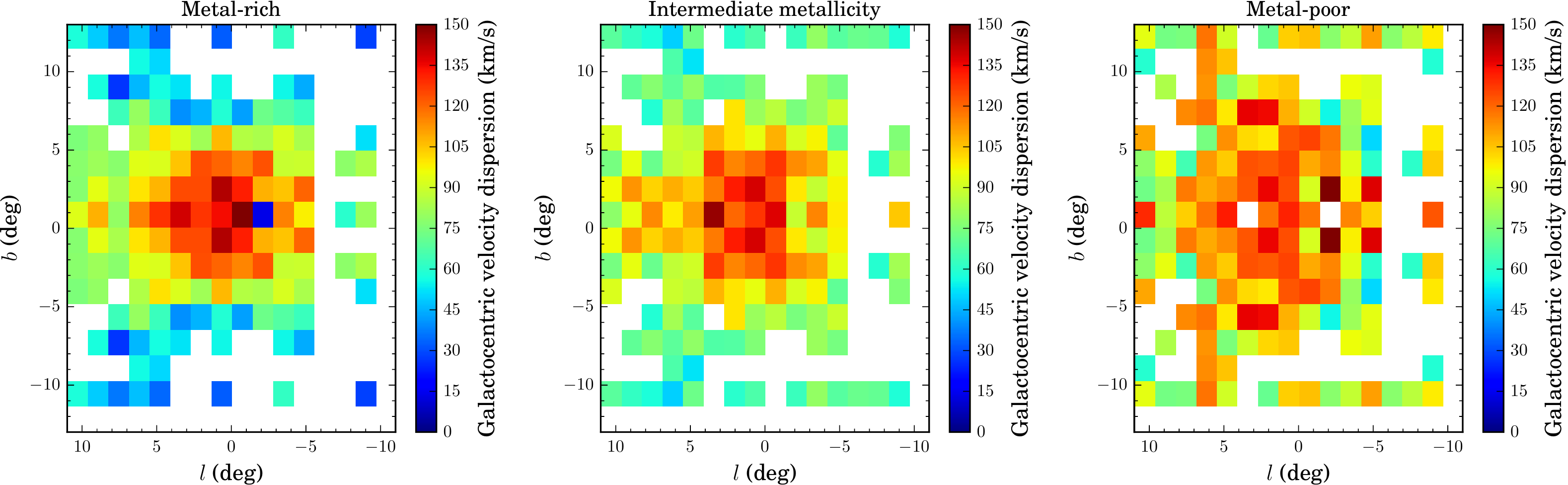}
\caption{$(l,b)$ kinematic maps for the three metallicity components found in the bulge MDF. In each case, the sample has been two-folded with respect to the Galactic plane, and the map has been constructed by binning into a $16\times16$ grid. Only pixels containing more than five stars are displayed. Top panels: mean Galactocentric velocity maps. Bottom panels: Galactocentric velocity dispersion maps.} 
\label{fig:lb_kin_maps_simplecut}
\end{figure*}

\begin{figure*}
\centering
\includegraphics[width=0.96\textwidth]{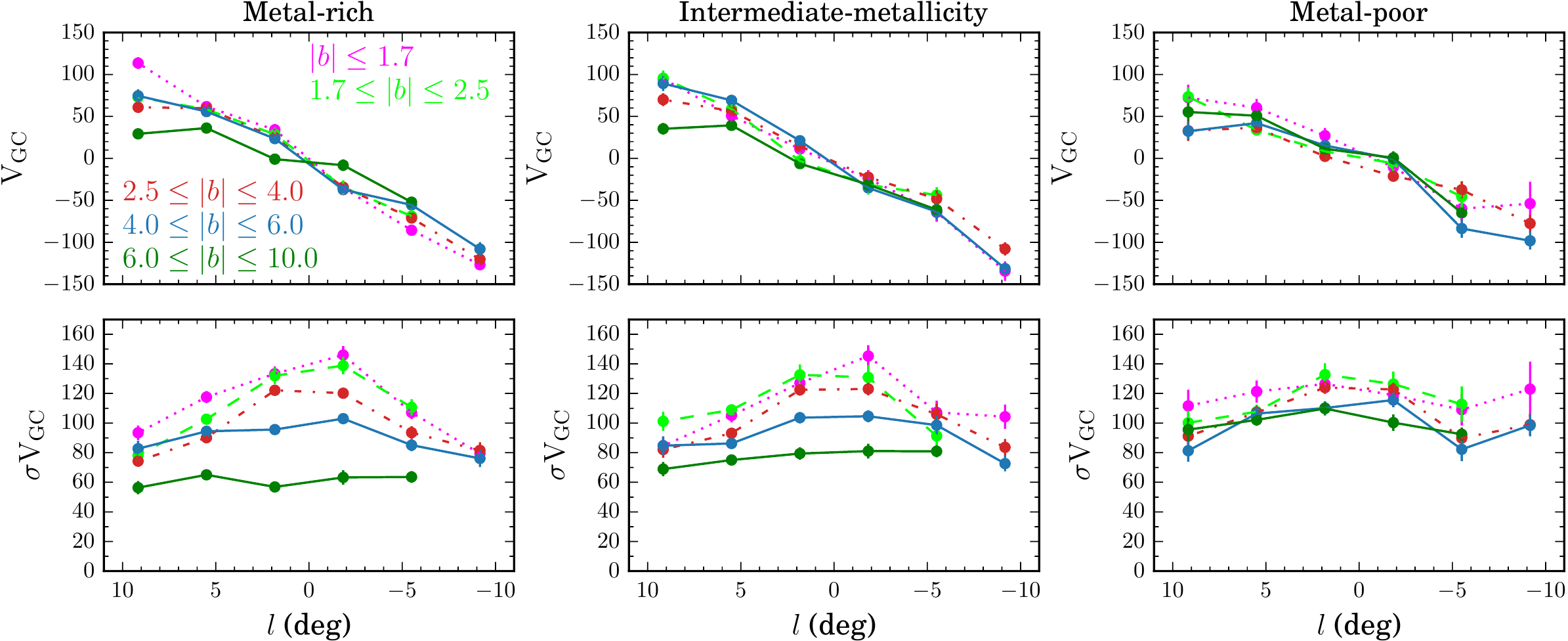}
\caption{Mean Galactocentric velocity and velocity dispersion curves for stars in the three GMM-inferred metallicity components. The color of the points and lines indicates the latitude range, using the same color scheme adopted elsewhere in this paper (e.g., Figs.~\ref{fig:lb_strips} and \ref{fig:mdf_strips}). }
\label{fig:lb_kin_curves_simplecut}
\end{figure*}

We examine in this section the kinematic patterns of stars most closely associated to each of the three GMM components. Figure~\ref{fig:lb_kin_maps_simplecut} contains maps of the mean velocity and velocity dispersion in the $(l,b)$ plane, computed from the bulge sample (folded about the midplane) in a grid of $(\Delta l, \Delta b) \sim (1.25^\circ, 1.5^\circ)$. As a complementary view, we display in Fig.\ref{fig:lb_kin_curves_simplecut} the mean and dispersion Galactocentric velocity curves for the three metallicity components separated into the latitude strips adopted elsewhere in this work.
In this general qualitative assessment, we attempt to construct subsamples of stars most likely to belong to each of the MDF components. To this end, we separate the  bulge sample into three metallicity ranges, centered on the three components inferred in Sect.~\ref{sec:mdf_gmm} but avoiding the regions of largest overlap (and so, where the association of a star with a given component is more uncertain). The sample is thus separated into the following ranges: $\rm [M/H]\geq+0.1$~dex (metal-rich), $\rm -0.4\leq[M/H]\leq0.0$~dex (metal-intermediate), and $\rm -1.2\leq[M/H]\leq-0.5$~dex (metal-poor).

From the upper set of panels of Figs.\ref{fig:lb_kin_maps_simplecut} and \ref{fig:lb_kin_curves_simplecut}, one can see that all three components show a cylindrical rotation pattern about the minor axis. The latitudinal variation of the rotation pattern (i.e. the departures from perfect cylindrical rotation) seem to be somehow larger for the metal-rich component compared to the other two. This component has in addition the most pronounced rotation in the midplane (compare for example curves for $|b|\leq1.7^\circ$). On the other hand a more uniform rotation pattern can be seen in stars belonging to the metal-poor component, which also show an overall slower rotation. Although the cylindrical rotation pattern in the bulge is known from previous optical spectroscopic surveys \citep{Zoccali14}, we characterize it here from stars distributed in a previously ill-explored region close to the midplane. Our results here update those previously found from APOGEE DR12 data \citep{Ness16,Zasowski16}, although with higher spatial resolution due to the larger sample available now.

The patterns of velocity dispersion (lower row of Fig.~\ref{fig:lb_kin_maps_simplecut} and \ref{fig:lb_kin_curves_simplecut}) show some differences between the metallicity components. In the metal-rich stars, the velocity dispersion has overall a large vertical variation, increasing sharply with decreasing latitude, with a pronounced enhancement in the central region within $|\ell|<4^\circ$ and $|b|<4^\circ$. 

This is also consistent with the results of \citet{Ness16,Zasowski16}. In contrast, Fig.~\ref{fig:lb_kin_maps_simplecut} and \ref{fig:lb_kin_curves_simplecut} show that the metal-intermediate component is overall kinematically hotter (less variation with latitude), and displays a less pronounced increment of velocity dispersion towards the center.
An even larger isotropy is seen for stars in the metal-poor group, which appears kinematically hot over nearly the entire sampled spatial region. This is clearly seen from both the map and the relative flat velocity dispersion curves.
As we discuss in Sect.~\ref{sec:discussion}, these varying projected kinematical patterns can be used to speculate on the orbital properties of the stars associated with each of the metallicity components.

\subsection{MDF Decomposition with Non-negative Matrix Factorization}
\label{sec:nmf}

The primary aim of this paper is to explore the complexity of the bulge's MDF using GMM (Sect.~\ref{sec:mdf_gmm}--\ref{subsec:kin_mdf_comps}), but in this section we describe an alternative decomposition using Non-negative Matrix Factorization (NMF; \citeauthor{Lee1999} \citeyear{Lee1999}, with some examples of astrophysical applications in \citeauthor{Igual08} \citeyear{Igual08} and \citeauthor{Hurley14} \citeyear{Hurley14}).
NMF is a dimensionality reduction technique similar to Principal Component Analysis (PCA) in spirit; one key difference is that the eigenvectors are constrained to be non-negative. This constraint makes NMF an appealing option for decomposing a spatially variant MDF, since in principle, the MDF at any location is a superposition of every MDF component scaled by a non-negative value (including 0). An additional benefit of NMF is that the components themselves are not limited to Gaussians or any pre-defined functional form.

\begin{figure*}
\centering
\includegraphics[width=0.96\textwidth]{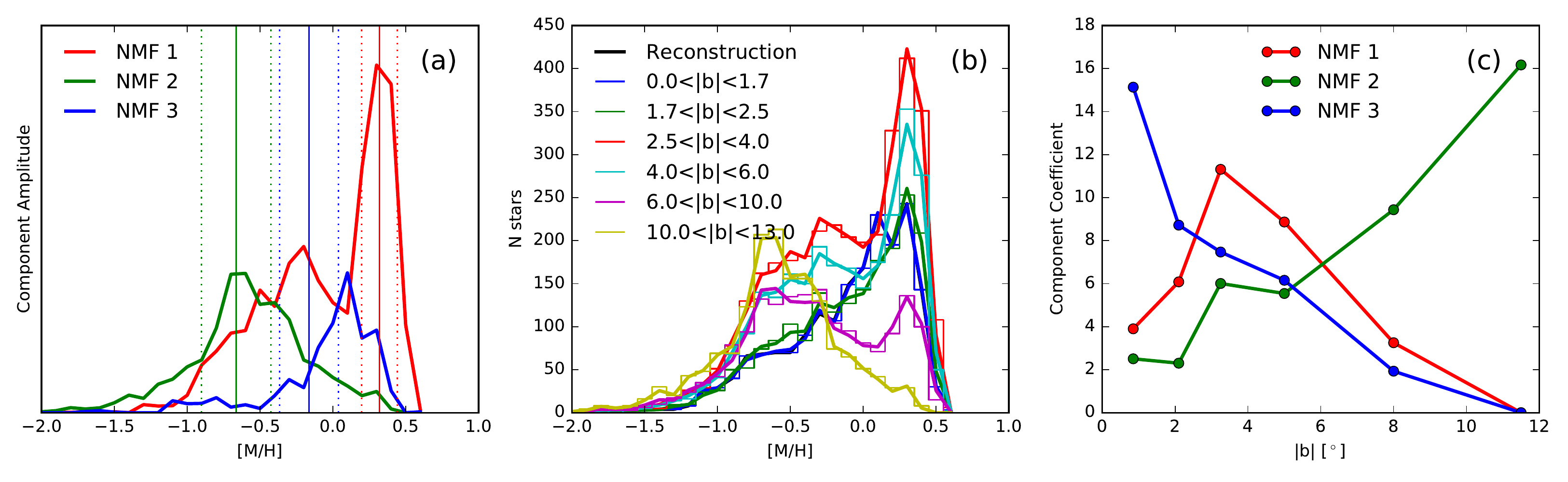}
\caption{
Non-negative matrix factorization (NMF) as applied to our bulge MDFs. 
Panel~(a): The components inferred from a three-component decomposition applied to the $|b|$-dependent MDFs discussed in Sect.~\ref{sec:mdf_properties} and shown in Fig.~\ref{fig:mdf_strips}. The vertical solid and dotted lines indicate the mean and $\pm\sigma$ of the three GMM components extracted in Sect.~\ref{sec:mdf_gmm}.
Panel~(b): Reconstruction of each MDF using the components in Panel~(a).
Panel~(c): Relative weight of the NMF components in each latitude strip.} 
\label{fig:nmf}
\end{figure*}

Using the same absolute latitude bins described above (Fig.~\ref{fig:lb_strips}), and motivated by the robustness of the three-component GMM solution, we perform a three-component NMF decomposition on the same MDFs over the range $\rm -2 \le [M/H] \le +0.7$~dex ($\rm \Delta[M/H]=0.1$~dex). The NMF components and the reconstructed MDFs are shown in the left and middle panels of Fig.~\ref{fig:nmf}, respectively.  When rerunning the decomposition while removing one of the input MDFs, the missing MDF is able to be reconstructed using the output components to the same apparent quality as when all MDFs are included.  We performed this same procedure on the subsampled MDFs described in Sect.~\ref{subsec:mdf_comp_other_surveys} and found no difference in the shape of the components, simply a reduced amplitude.

In comparing the NMF components to the range of GMM components (vertical lines in Fig.~\ref{fig:nmf}a), we find broad consistency along with some marked differences, some of which are simply due to the different methodologies and some of which may contain meaningful information.  Regardless of the weights given to the three components, a trimodal distribution results, in agreement with the GMM findings\footnote{The exception is if a zero weight is given to the broadest, most metal-rich eigenvector and not to the other two, which does not match any of the patterns in our dataset.}. The metal-poorest NMF component (NMF~2, green line) peaks at the same metallicity as the metal-poor GMM component, with a large width. The metal-richest NMF component (NMF~1, red line) peaks at the same metallicity as the metal-rich GMM component, with a small width in that peak at $\rm [M/H] \sim +0.3$~dex; however, the metal-richest NMF component is the widest of the three, and significantly, shows a secondary peak that coincides with the metallicity of the metal-intermediate GMM component.  

These two peaks in the eigenvector (or ``eigenMDF'') imply that the numbers of stars at these metallicities are somewhat correlated with each other across the range of $|b|$ spanned by the MDFs.  An alternative way of interpreting this is: in contrast to representing the MDF by summed Gaussians with smoothly varying relative heights, the MDF can be modeled equally well with a skewed, multi-peaked distribution that is modulated by a second (narrower, symmetric, single-peaked) component (NMF~3, blue line in Figure~\ref{fig:nmf}a, centered at $\rm [M/H] \sim 0$~dex).  The near-total dominance of NMF~2 at high $|b|$ suggests that it may represent the combined distributions of inner halo and thick disc stars, which contribute relatively little at low $|Z|$; the inner bulge MDF variations are driven by the relative strength of the skewed, multi-peaked distribution and the symmetric, solar-metallicity distribution.

The details of this comparison depend only weakly on our choice of a three-component decomposition. If a different number of components is chosen, each one's shape changes slightly, but the general pattern is robust. For example, in a two-component decomposition, the dominant NMF~1 has a shoulder instead of a dip at $\rm [M/H] \sim 0$~dex, while NMF~2 is nearly identical. In a four- or five-component decomposition, the three eigenMDFs shown in Figure~\ref{fig:nmf}a are again nearly identical, with additional components providing only slight modulations with very low weights.

Our goal in exploring this technique is to demonstrate an alternative approach to MDF analysis, as a contrast to adopting Gaussian (or similar) bases and identifying or classifying ``populations'' to associate with known or assumed bulge constituents. These types of non-parametric decompositions may be useful when comparing to, e.g., stellar abundance distributions produced by chemical enrichment models with an extended, complex star formation history, which do not predict distributions easily modeled by a small number of symmetric Gaussians. A deeper exploration of the full information contained in a NMF decomposition (including the optimal number of components) and in other techniques is deferred to future work.

\section{Discussion}
\label{sec:discussion}

\subsection{MDF complexity from the perspective of other surveys}
\label{subsec:mdf_comp_other_surveys}

In Sect.~\ref{sec:mdf_gmm} we studied the varying structure of the bulge MDF in several bins sampling different distances from the Galactic midplane. 
In this section, we compare this picture with the general conclusions drawn from the results of optical spectroscopic surveys in the bulge region, some of which found a larger optimal number of components and others which found a smaller number.

A complex picture, with a higher number of components, emerged as a main outcome of the ARGOS survey. The primary targets were bulge red clump (RC) stars that were observed in the calcium triplet (CaT) region at $R=\lambda/\Delta\lambda\sim11,000$. A full spectrum-fitting approach was used to estimate fundamental parameters, including metallicity (but not $\rm T_{eff}$, which was fixed from photometric calibrations), with an uncertainty of $\sim$0.1~dex. The structure of the MDF was studied from three latitude bins at $b=-5^\circ$, $-7.5^\circ$, and $-10^\circ$ (within $l\pm15^\circ$), each containing between 2000 and 4000 stars. From these data, \citet{Ness13} found the MDF shape to be optimally decomposed by up to five Gaussian components, lettered A through E in order of decreasing metallicity, with the three most metal-rich components (at $\rm [Fe/H]=+0.1$, $-0.28$, and $-0.68$~dex) dominating the density structure of the MDF. An overall comparable picture was reached from the analysis of the whole sample of microlensed bulge dwarf stars of \citet{Bensby17}. Their MDF, based on a small sample of 90 stars but observed at high resolution ($R=42,000-48,000$), suggested the presence of up to five peaks, at similar positions to those of ARGOS.

Our results (Fig.~\ref{fig:mdf_strips}) are comparable with those of ARGOS, at least with regards to the number of Gaussian components needed to reproduce the observed MDF at ${\rm[Fe/H]}\geq-1$~dex. Our sample has too few stars with ${\rm [Fe/H]\leq-1}$~dex to be represented by extra components in the Gaussian mixture. Thus, the larger size of the ARGOS sample may explain the presence of two additional minor components at $\rm [Fe/H] \sim-1.2$~dex and $\rm [Fe/H]\sim-1.7$~dex. Although the number of components at $\rm [Fe/H]\geq-1$~dex is the same, the centers of our metallicity components ($-0.66$, $-0.17$ and $+0.32$~dex) are not consistent with those of ARGOS, nor are they simply shifted due to a constant offset between the survey metallicity scales. For example, the separation between our metal-rich and metal-intermediate components is 0.10~dex larger than the separation between components A and B of ARGOS (which is statistically significant, given the errorbars of the respective centroid estimates in both datasets).

A similar multimodal MDF decomposition was obtained from a sample of $\sim7500$ stars from APOGEE DR12 \citep{garciaperez18}. A three Gaussian decomposition was evaluated in a number of mid-size samples ($\sim200$ stars in average) separating the bulge area in broad spatial ($R_{GC}$, $Z$) pixels \citep[see Fig. 9 of][]{garciaperez18}. As a result, four different metallicities ($+0.32$, $+0.0$, $-0.46$ and $-0.83$~dex) were found as the centroid of the components considering the results of all fields together (i. e. not every component was detected in all fields). As in the case of ARGOS, the position of the components seem inconsistent with the results obtained here. Note that the results presented in this work, as based on a more recent and improved APOGEE data release, and on a larger number of stars, supersede those of \citep{garciaperez18}.

In contrast, \citet{Hill11} used a GMM analysis on 219 RC stars to investigate the shape of the MDF in Baade's Window and found evidence for only two populations, of roughly equal size, located at ${\rm [Fe/H]=-0.3}$~dex and $+0.3$~dex. A bimodal MDF was also posited to describe a large area below the midplane, based on results from the GIBS and GES surveys \citep{Gonzalez15,Zoccali17,Rojas-Arriagada14,Rojas-Arriagada17}. As optical surveys --- both of them observing with FLAMES@VLT --- the footprints of these observations were concentrated in the area below $b\lesssim-3.5^\circ$ (although GIBS also had a strip of fields at $b=-2^\circ$). As in ARGOS, the targets were RC giants observed in the region around the CaT. In the case of GIBS, spectra were obtained at $R\sim6500$, and a CaT calibration was used to obtain metallicities with a nominal uncertainty of $0.2$~dex \citep{Vasquez15}. GES data were observed at higher spectral resolution ($R\sim17,000$) and analyzed through a full spectrum fitting analysis, reaching a metallicity uncertainty of $\sim0.1-0.15$~dex. In both surveys, the typical sample size per observed field is around a couple hundred stars. The overall picture, consistent between \citet{Hill11}, GES, and GIBS, is of a bimodal bulge MDF. The metallicity positions of the two components vary slightly between the different fields in each study, but are found to be located at about $-0.4\ /\ +0.3$~dex (GIBS) and $-0.35\ /\ +0.40$~dex (GES), compared to $\pm 0.3$ in \citet{Hill11}.

In a different approach, \citet{Fragkoudi18} compared the shape of the MDF obtained from APOGEE DR13 with a N-body simulation in a number of ($l$,$b$) pixels. The simulation tracked the secular evolution of a thin disc and two thick discs into a B/P bulge. From the model, simulated  MDFs were ``observed'' by reproducing the line-of-sight distance sampling of APOGEE. These simulated MDFs were found to qualitatively agree with the data in the sense of being visually bimodal; they also reproduced the spatial variation of the mean metallicity in the bulge region (vertical and longitudinal gradients) and the variation in the relative contribution of particles from the different discs in each observed field.

In spite of the simpler ($N=2$) decomposition found by these studies to explain the shape of the MDF, the optically derived distributions are in good qualitative agreement with those in our latitude strips spanning similar areas. In fact, the position of our metal-rich component is between the metal-rich components found by GIBS and GES. The shape of the MDF in our latitude bins spanning areas similar to that sampled by GIBS/GES can be visually described as a bimodal distribution (Sect.~\ref{sec:mdf_properties}).  

\begin{figure}
\centering
\includegraphics[width=0.48\textwidth]{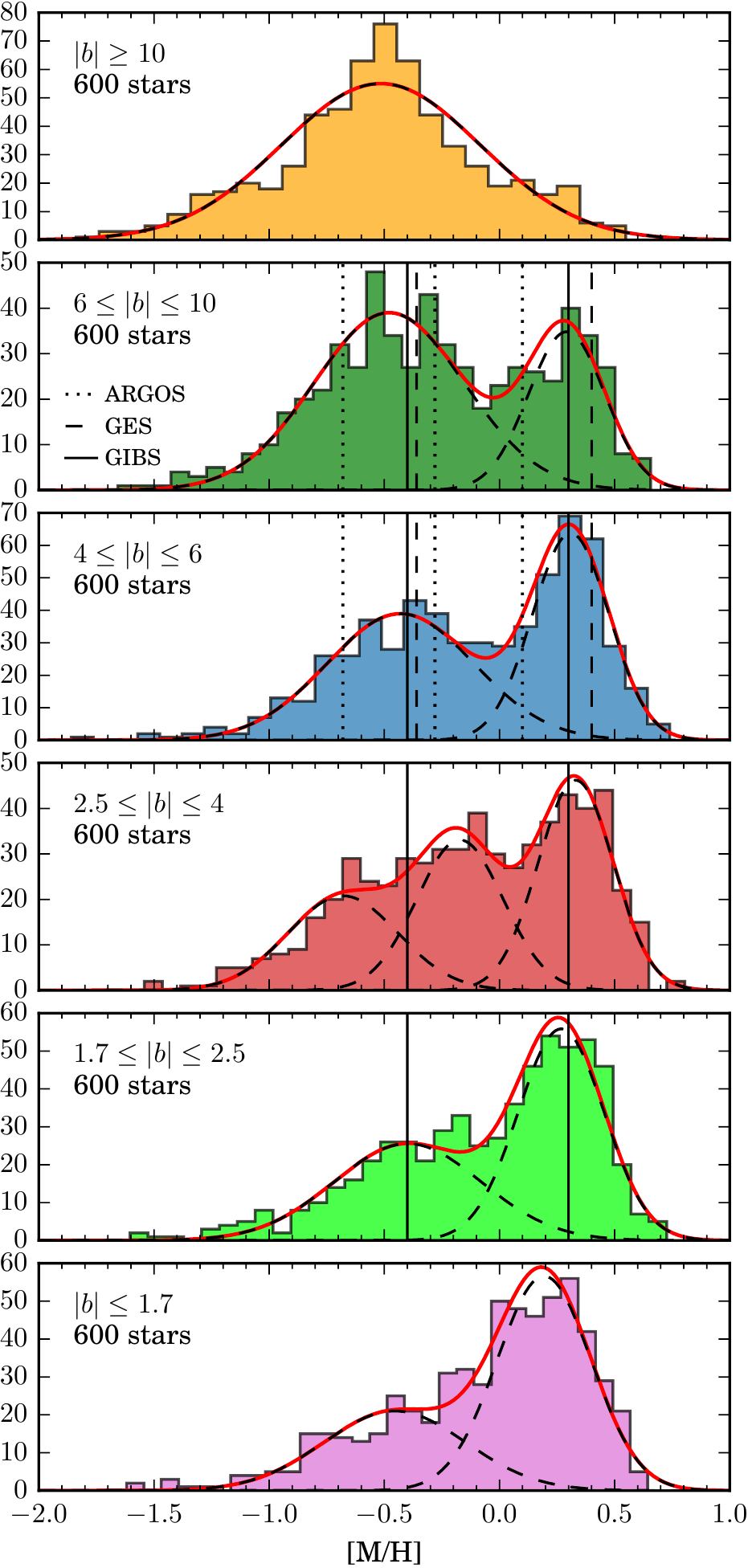}
\caption{Gaussian mixture decomposition of MDFs designed to reproduce the sampling of previous spectroscopic surveys (analogous to Fig.~\ref{fig:mdf_strips}). From each latitude sample, a subsample of 600 stars was taken and their individual measurement errors inflated to $0.1$~dex. The best-fit GMM mixture to each of these resulting samples is overplotted with dashed black and red solid lines, depicting individual components and the total mixture, respectively. The vertical black lines indicate the position of the peaks as found from ARGOS (dotted), GES (dashed) and GIBS (solid) MDF decompositions. They are only displayed in the panels approximately corresponding to the latitude ranges they cover.}
\label{fig:mdf_strips_simOpticalSurveys}
\end{figure}

Quantitatively, however, this lower-metallicity half of the bimodal distribution is better described in our dataset by two Gaussian components, which we posit are unresolved when the sample sizes are small and/or the individual metallicity uncertainties are large.
Thanks to the large samples we have in each of our latitude strips, and the small individual measurement errors, we can perform an experiment simulating previous optical surveys' sampling of the bulge MDF, which is illustrated in Fig.~\ref{fig:mdf_strips_simOpticalSurveys}. We downsampled our six latitude strips by randomly choosing 600 stars from each of them. Although this sample size is a considerable reduction from the current sample size, it is still larger than that in the typical GIBS/GES fields ($\sim200$). Gaussian noise was added to the individual metallicities to simulate inflating their uncertainties to a $\sim0.12$~dex level.

The optimal GMM solution was computed for each distribution allowing for one to six components and adopting the BIC as model selection criterion. The optimal mixtures fitted to each MDF are displayed on top of them in Fig.~\ref{fig:mdf_strips_simOpticalSurveys}. As we can see, the MDF of the outermost bin at $|b|\geq10^\circ$ is now found to be well represented by a single Gaussian component centered at $\sim-0.5$~dex. On the other hand, the MDF of the bins inside $|b|=10^\circ$ are now well represented by bimodal distributions which in every case look as a fair representation of the observed density distribution. In one case, the $2.5^\circ\leq|b|\leq4^\circ$ strip, the solution is still found to be trimodal and consistent with the result obtained before from the full original sample. In fact, the whole exercise was run several times, and in some cases this distribution was well represented by a bimodal mixture. We choose to leave this instance of the experiment to show that with smaller samples with larger individual errors the results of a GMM run become simpler, or in any case, less robust, since the observed GMM does not provide enough information to be explained by a more complex Gaussian mixture.

Overall, Fig.~\ref{fig:mdf_strips_simOpticalSurveys} shows that under similar sampling conditions, the qualitative shape of the MDF in fields over the same spatial area is quite consistent between APOGEE and previous optical surveys. In all cases, the metal-rich component, apparent by simple visual inspection, is appropriately detected and described by a component in the Gaussian mixture. On the other hand, only a larger and more precise sampling allows to find extra complexity at the sub-solar metallicity domain of the MDF.

The metallicity distribution in a given bulge line of sight results from the varying contributions of what could eventually be several structures coexisting in the inner Galaxy. If any pair of them are characterized by intrinsically similar distributions with a significant overlap, they might become difficult to disentangle from the observed MDF. The situation can be even more complex if the intrinsic shape of the distributions depart from the simple assumed Gaussianity (e.g., Sect.~\ref{sec:nmf}).

To summarize, the need for three Gaussian components to explain the shape of the bulge MDF qualitatively agrees with the results from the ARGOS survey. It is also in agreement with the density structure reported from other large optical spectroscopic surveys once considerations of sample size and individual error measurements are taken into account. Whether these components represent underlying physical components or not can be assessed from the examination of the distributions of other observed properties.

\subsection{The bulge's MDF in the context of bulge formation mechanisms and extragalactic evidence}
\label{subsec:mdf_comments}

Our trimodal characterization of the bulge MDF reveals different spatial variations and kinematical correlations for stars in the three metallicity groups. These observations are evidence for understanding the metallicity components as distinct physical populations \citep[see also][]{Queiroz20}. It is interesting to review these findings in the context of extragalactic observations and theoretical predictions from bulge formation scenarios. We will start with the highest-metallicity stars.

Many lines of observational evidence have been used to argue that the bulge's metal-rich stars are entrained in the B/P  X-shaped bar \citep[e.g.,][]{Ness12,Rojas-Arriagada17,Zoccali17,Bensby17}.  
In the canonical prescription for bar formation, secular evolution proceeds from spontaneous gravitational instabilities that rearrange angular momentum in the pre-existing disc and result in mass transfer, building a stellar concentration in a central bar. Subsequent bending instabilities produce a B/P bulge from the stars being kinematically heated in the vertical direction. The point here is that secular evolution is fundamentally a dissipationless process taking place in a gas-poor phase of galactic evolution, driven by stellar dynamics. This process has implications for the ages and present-day kinematics of stars entrained in the bar.

With respect to age, one must be careful to distinguish between ages of the bar stars and the age of the bar structure itself. The high metallicity of the stars argued to be in the bar may imply younger ages, relative to the rest of the bulge, which may suggest that the bar itself is a relatively recent addition to the MW (\citeauthor{Dimatteo16} \citeyear{Dimatteo16}, \citeauthor{Fragkoudi18} \citeyear{Fragkoudi18}, \citeauthor{Wegg19} \citeyear{Wegg19}, Lian et al. in prep, but for a different view, \citeauthor{Bovy19} \citeyear{Bovy19}).  It has been established from the study of the redshift-dependent fraction of galactic bars, in observations \citep{Sheth08} and cosmological simulations \citep{Kraljic12, Fragkoudi20}, that bars appear at lookback times of $\sim$8~Gyr in MW-mass galaxies---that is, at a later time compared with the main epoch of mass assembly and star formation. However, it has also been shown that the presence of a rotating non-axisymmetric bar potential can induce a loss of angular momentum in gas clouds, which fall to the center, trigger star formation \citep{Ellison11}, and produce stars the same age or younger than the initial bar itself. Thus the ages of individual stars constrained, by other means, to be part of the bar population are essential for disentangling the formation history of not only the bar's stars but also the bar's structure. Detections of a significant fraction of relatively young stars associated with the bar are important lines of evidence \citep[e.g. ][]{Bensby11,Bensby13,Bensby17,Schultheis17,Hasselquist20}. 

The kinematics of our metal-richest stars support the interpretation of bar membership. The cylindrical rotation pattern exhibited by metal-rich stars in Figs.~\ref{fig:lb_kin_maps_simplecut}--\ref{fig:lb_kin_curves_simplecut} is an observational outcome predicted by $N$-body simulations of B/P bulges formed from the secular evolution of the disc through bar formation and buckling \citep{Zhao96}. It is also a phenomenology observed in external local galaxies with central B/P bar-dominated morphology \citep{Molaeinezhad16}. The slightly enhanced rotation pattern towards the midplane observed in our data represents an additional signature of bar-dominated dynamics \citep{Gomez18}.
The $\sigma(V_{GC})$ maps (Fig.~\ref{fig:lb_kin_maps_simplecut}) show that this component has a steep vertical kinematical gradient, consistent with the progressive dominance of bar streaming orbits towards the midplane. At the same time, such an orbital structure may explain the central $\sigma(V_{GC})$ enhancement (at $|\ell|\leq4^\circ$, $|b|\leq4^\circ$) as an effect of the lines of sight crossing inner regions where the several families of orbits supporting the bar structure intersect each other.

Let us now turn our attention to the group of metal-intermediate bulge stars. This component, in our analysis, comprises most of the stars identified as belonging to the metal-poor component in optical studies such as GES/GIBS (Sect.~\ref{subsec:mdf_comp_other_surveys}), so that comparisons between them are pertinent in what follows.
As shown in Figs.~\ref{fig:lb_kin_maps_simplecut}--\ref{fig:lb_kin_curves_simplecut}, this component is overall kinematically hotter than the metal-rich one but has a less pronounced enhancement in dispersion towards the center. However, it displays a more coherent cylindrical rotation than the metal-rich stars. 
Can these stars be associated with the buckling instability of stars formed in the disc, or to a central in-situ formation scenario instead?

Stars in this metallicity range ($-0.5$ to $0.0$~dex) have been interpreted as the bulk of the early thin disc rearranged into the B/P bulge by the bar instability processes \citep{Ness13}. In this sense, the splitting of the red clump luminosity distribution (as a proxy of distance) was taken by \citet{Ness12} as evidence of them participating in the X-shape. However, in the same metallicity range, GES data suggest a less structured, more homogeneous spatial distribution \citep{Rojas-Arriagada17}. In addition, such a scenario conflicts with the relatively high levels of $\alpha$-element enhancement observed for these stars in the APOGEE data (in the same $R_{GC}\leq3.5$~kpc sampled in this work; \citeauthor{Rojas-Arriagada19} \citeyear{Rojas-Arriagada19}, \citeauthor{Queiroz_2020_starhorse} \citeyear{Queiroz_2020_starhorse}); this enhancement is instead indicative of a rapid star forming episode \citep[i.e., the chemical evolution models of][]{Rojas-Arriagada17, Matteucci20, Lian2020}.

Using simulations, \citet{Debattista17} argue that the bar could kinematically separate co-spatial populations according to their initial in-plane radial random motions. This {\it kinematical fractionation} determines the height stars can reach during the buckling bar event, with those being initially kinematically cooler forming a stronger X-shaped spatial distribution, while hotter populations ending instead in a weaker bar vertically thicker box. As older stars have hotter initial kinematics and different chemistry, this scenario implies a continuum of current bulge stellar content properties as a function of metallicity. In particular, this scenario can reconcile the the weaker (or absent) RC luminosity distribution splitting, and the particular alpha-enhancement of stars over the metallicity range of our metal-intermediate component, given an appropriate chemo-dynamical structure of the initial disk(s).

In a different proposal, \citet{Dimatteo16} and \citet{Fragkoudi18} associate these metal-intermediate stars with the larger thick disc in order to reproduce their observed chemical and kinematical distributions. Although a qualitatively similar chemistry is found for the sequences of bulge and thick disc stars in the ${\rm [\alpha/Fe]}$~vs~[Fe/H] plane, a more metal-rich position has been found for the bulge ``knee'' 
(\citeauthor{Cunha06} \citeyear{Cunha06}, \citeauthor{Rojas-Arriagada17} \citeyear{Rojas-Arriagada17}, although this difference might be smaller if comparing with the inner thick disk; \citeauthor{Zasowski2019} \citeyear{Zasowski2019}),
indicating a higher early star formation rate in the bulge than in the thick disc inside the solar circle. This has been also suggested to be the case from the comparison of microlensed bulge dwarf stars with local thick disk dwarfs \citep{Bensby13,Bensby17}. In addition, these works suggest that bulge stars might be overall slightly more alpha-enhanced than thick disk stars. This latter result has been confirmed in \citet{Queiroz20} by comparing the bulge and inner thick disk ($|X|<5$~kpc, $|Y|<3.5$~kpc and $|Z|<1$~kpc) chemical distributions.

Alternatively, these stars have been proposed as a morphologically classical/spheroid bulge component \citep{Babusiaux10,Hill11,Rojas-Arriagada17}. In $\Lambda$-CDM, this component corresponds to mass assembled from the dissipationless merging of stellar substructures. There are a couple of observational constraints limiting the plausibility of this scenario: (i) a bulge assembled from many different building blocks might lead to a rather chemically inhomogeneous structure, unless the independent building blocks were already uniform to begin with. What is observed, however, is a tight, seemingly singular chemical evolutionary sequence, at least in the alpha elements \citep{Zasowski2019,Rojas-Arriagada19}. (ii) This hypothesis would require a rather high mass for the individual building blocks. Indeed, if their individual potential wells were shallow, the consequent slow SFR would be unable to produce a substantial amount of stars of a high enough metallicity to match our component distribution.

Moreover, the analysis of 30 high resolution hydrodynamical cosmological simulations have shown that the amount of accreted stars in the inner/bulge region of MW-mass galaxies is very low, even negligible in some cases. Indeed, most of the (already low) accreted components of these simulated bulges come from a very small number of building blocks \citep{Gargiulo19}. This result is confirmed by the analysis of zoom-in simulations of \citep{Fragkoudi20}, which has revealed that the modeled galaxies that most closely resemble the chemodynamical properties of the Milky Way are characterized by a quiescent merger history: the last major merger happening long before bar formation, and overall, with a negligible fraction of ex-situ stars in the bulge region.

Finally, an in-situ formation scenario may provide an appropriate explanation for the present-day observed properties of metal-intermediate stars. Observations at high redshift ($z=1.5-2.2$) have revealed the presence of central stellar overdensities in galaxies that otherwise are still actively forming stars in their gas-rich discs in an inside-out way \citep{Tacchella15,Nelson16}. This implies that mature bulges are already in place early during the main epoch of star formation. 
This mechanism rapidly forms rotating bulges out of the gas-rich disc in a internal dissipative process, without the need for major mergers \citep{Tadaki17}. Theoretical prescriptions of such a scenario include the formation of giant disc gas clumps, their migration and coalescence \citep{Immeli04,Elmegreen08} at the center, or global violent disc instabilities leading to a massive central accumulation of star-forming gas \citep{Dekel14}.
In addition, in the aforementioned study of cosmological simulations by \citet{Gargiulo19}, they found that in 75\% of the cases most of the stars found in the bulge at present day formed in-situ in the inner regions of the simulated galaxies.
A fast and early vigorous star formation in gas-rich high-redshift environments can produce a bulge with the properties observed in our metal-intermediate stars: a spatially extended, rotating, kinematically hot, relatively metal-rich and alpha enhanced stellar structure composed predominantly by old stars. In this sense, our metal-intermediate stars may correspond to the present day structure assembled centrally in-situ and early in the formation of the Milky Way. 

Our metal-poor component amounts for a relatively constant and minor fraction of the stars below solar metallicity. They display a tight but slow cylindrical rotation pattern and are kinematically hot and isotropic over most of the studied area. In \citet{Ness13} these stars were associated with the thick disc present in the inner Galaxy before the instability event. This interpretation conflicts with the peak of this component being located at ${\rm [M/H]=-0.66}$~dex, which is around 0.3/0.4~dex lower than the peak of the thick disc at ${\rm 3\leq R_{GC}\leq5}$~kpc \citep{hayden15,Queiroz_2020_starhorse}. On the other hand, they are too metal-rich to be associated with the inner halo. Nonetheless, our sample contains only a minor fraction of stars with ${\rm [M/H]\leq-1}$~dex, and therefore we cannot rule out that some of these stars belong to the stellar halo. In fact, from the early evidence of \citet{Minniti96} it is known that giant stars more metal poor than ${\rm [M/H]=-1}$~dex display halo-like kinematics, and should be naturally found in the inner Galactic regions but as an overall very small fraction.

In that metal-poor metallicity range, \citet{Kunder20} has proposed the existence of two different components traced by the intrinsically metal-poor old RR Lyrae stars, with the most centrally concentrated one proposed to be a classical bulge in the sense of being produced by an accretion event at high redshift. Whether our metal-poor stars can be associated with the stellar populations being traced by RR Lyrae at ${\rm [Fe/H]\sim-1}$~dex remains to be proven. Nonetheless, a merger scenario for this component has to account, as in the case of metal-intermediate stars, for their low dispersion of $\alpha$-elements. The latter also challenges a potential interpretation of this component as the product of accreted globular cluster stars; not only the chemical homogeneity of $\alpha$-elements is not guaranteed, but also the peak in metallicity of this component is in between the two peaks of the metallicity distribution of bulge globular clusters \citep[][their Fig. 4]{Bica16}. Field stars with chemical anomalies have been reported in this metallicity regime \citep{Schiavon17,Fernandez-Trincado19}, but they are estimated to amount for $\sim2\%$ of the total budget mass. The more recent work of \citep{Horta20} estimates that up to 1/3 of all bulge metal-poor stars (with ${\rm [Fe/H]\leq-0.8}$~dex) might be accreted, as identified from their chemical composition resembling those of low mass satellites of the MW.
Further data will be needed to resolve the formation scenario for these stars.

\section{Summary}
\label{sec:summary}

We have investigated the shape of the bulge MDF from a sample of $\sim13,000$ giant stars spatially located in the bulge region coming from the APOGEE DR16 and further internal incremental data releases. This sample allowed us to study the inner-Galaxy stellar populations over the whole $|\ell|\leq11^\circ$, $|b|\leq13^\circ$ area, and down to the midplane where the high levels of extinction have hindered previous attempts from optical surveys.

\noindent Our main findings are as follows:
\begin{itemize}
\item[(i)] The effects of the APOGEE selection function on the shape of the MDF is rather small and does not impact the complex density structure revealed by our analysis.

\item[(ii)] The ${\rm R_{GC}=3.5}$~kpc limit often adopted to spatially separate bulge/disc stellar samples finds qualitative justification from the changes in kinematical patterns characterized from our data around that limit.

\item[(iii)] The shape and spatial variations of the bulge MDF can be explained as the varying contribution of three main components at ${\rm [Fe/H]\geq-1}$~dex.

\item[(iv)] Our GMM analysis and robustness assessment show that the trimodal Gaussian mixture correctly captures the density structure of the MDF over the whole set of latitude strips we adopted to separate the bulge sample. We estimate that bulge populations dominate the line-of-sight samples inside $|b|=8^\circ$. Over that region, the mean metallicity of the components remain rather constant, and the widths present only mild variations. The relative weights of the metal-rich and metal-intermediate components determine the vertical variations of the density shape of the MDF, and so, naturally explain the vertical variations of summary quantities such as the metallicity gradient.

\item[(v)] In the same vein, we show that the bimodal Gaussian mixture found to well represent the MDF from optical surveys data could be an effect of the combination of relatively small sampling and larger individual measurement errors.

\item[(vi)] The kinematics of our three metallicity components show some differences which we relate to their different orbital structure.

\item[(vii)] Finally, we cast our results in the context of previous works on the bulge MDF, as well as of evidences from the study of high redshift galaxies and theoretical formation models. 
\end{itemize}

With the present data we are not in position to settle down a definitive physical interpretation of the nature and origin of the bulge components. Nonetheless, our data allowed us to trace the big empirical picture, conciliating previous observational evidence and contributing to assemble a state-of-the-art observational picture of the bulge stellar populations down to the midplane. The complexities revealed here and in other studies highlight the need of considering both local (resolved stellar populations) and high redshift (integrated properties) evidence to piece together the sequence of events behind galaxy formation and evolution. This observational approach may lead us to assemble, with the help of theoretical models/prescriptions, a scenario providing a satisfactory and harmonious account of galaxy phenomenology from $z=2$ to the present-day picture offered by the Milky Way.


\section*{Acknowledgements}
ARA and GZ are grateful to the Observatoire de la C\^{o}te d'Azur for hospitality during visits that greatly benefited the analysis in this paper. ARA acknowledges support from FONDECYT through grant 3180203. The authors thank R.~M.~Rich for his compelling analogy between bulge metallicity components and martinis during the 2018 ESO bulge meeting in Puc\'{o}n, which served as an inspiration for this paper. D.G. gratefully acknowledges support from the Chilean Centro de Excelencia en Astrof\'isica y Tecnolog\'ias Afines (CATA) BASAL grant AFB-170002. D.G. and D.M. gratefully acknowledge support from the Chilean Centro de Excelencia en Astrof\'isica y Tecnolog\'ias Afines (CATA) BASAL grant AFB-170002. D.M. also thanks support from proyecto Fondecyt Regular No. 1170121. D.G. acknowledges financial support from the Direcci\'on de Investigaci\'on y Desarrollo de la Universidad de La Serena through the Programa de Incentivo a la Investigaci\'on de Acad\'emicos (PIA-DIDULS). DAGH acknowledges support from the State Research Agency (AEI) of the Spanish Ministry of Science, Innovation and Universities (MCIU) and the European Regional Development Fund (FEDER) under grant AYA2017-88254-P. J.G.F-T is supported by FONDECYT No. 3180210 and Becas Iberoam\'erica Investigador 2019, Banco Santander Chile. AM acknowledges financial support from FONDECYT Regular 1181797 and funding from the Max Planck Society through a Partner Group grant. CC acknowledges partial support from DFG Grant CH1188/2-1 and from the ChETEC COST Action (CA16117), supported by COST (European Cooperation in Science and Technology).\\

This work has made use of data from the European Space Agency (ESA) mission {\it Gaia} (\url{https://www.cosmos.esa.int/gaia}), processed by the {\it Gaia} Data Processing and Analysis Consortium (DPAC, \url{https://www.cosmos.esa.int/web/gaia/dpac/consortium}). Funding for the DPAC has been provided by national institutions, in particular the institutions participating in the {\it Gaia} Multilateral Agreement.\\

Funding for the Sloan Digital Sky Survey IV has been provided by the Alfred P. Sloan Foundation, the U.S. Department of Energy Office of Science, and the Participating Institutions. SDSS-IV acknowledges support and resources from the Center for High-Performance Computing at
the University of Utah. The SDSS web site is www.sdss.org. \\

SDSS-IV is managed by the Astrophysical Research Consortium for the Participating Institutions of the SDSS Collaboration including the Brazilian Participation Group, the Carnegie Institution for Science, Carnegie Mellon University, the Chilean Participation Group, the French Participation Group, Harvard-Smithsonian Center for Astrophysics, Instituto de Astrof\'isica de Canarias, The Johns Hopkins University, Kavli Institute for the Physics and Mathematics of the Universe (IPMU) / University of Tokyo, the Korean Participation Group, Lawrence Berkeley National Laboratory, Leibniz Institut f\"ur Astrophysik Potsdam (AIP),  Max-Planck-Institut f\"ur Astronomie (MPIA Heidelberg), Max-Planck-Institut f\"ur Astrophysik (MPA Garching), Max-Planck-Institut f\"ur Extraterrestrische Physik (MPE), National Astronomical Observatories of China, New Mexico State University, New York University, University of Notre Dame, Observat\'ario Nacional / MCTI, The Ohio State University, Pennsylvania State University, Shanghai Astronomical Observatory, United Kingdom Participation Group, Universidad Nacional Aut\'onoma de M\'exico, University of Arizona, University of Colorado Boulder, University of Oxford, University of Portsmouth, University of Utah, University of Virginia, University of Washington, University of Wisconsin, Vanderbilt University, and Yale University.

\section*{Data Availability}

The data used in this article is from an internal incremental release of the SDSS-IV/APOGEE survey, following the SDSS-IV public Data Release 16. This means that a minor fraction of the sample studied here is not publicly available now but will be included in the final public data release of SDSS-IV.

\bibliographystyle{mnras}
\bibliography{MDF} 


\appendix

\section{Distance Validation}
\label{ap:sec:distance_validation}

\begin{figure*}
\begin{center}
\includegraphics[angle=0,trim=0in 0in 0in 0in, clip, width=0.95\textwidth]{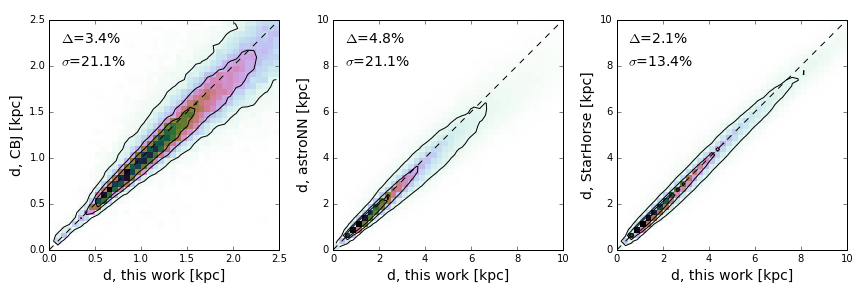} 
\caption{
Comparison of the spectro-photometric distances computed in this work (Sect.~\ref{subsec:dists_orbits}) with three others: {\it Gaia} Bayesian distances from \citet{Bailer-Jones18}, the neural network {\sc astroNN} distances from \citet{Leung19}, and  the spectrophotometric distances computed with the {\sc StarHorse} code \citep{Queiroz18}. In each case, we display the direct comparison of our spectrophotometric distances with those of the different catalogs for the APOGEE DR16 sources in common. The dashed black line indicates the 1:1 relation, and the contours trace the 50\%, 25\%, and 5\% isodensity lines. We find generally good agreement with all three methods.}
\label{fig:distances_methods}
\end{center}
\end{figure*}

\begin{figure*}
\begin{center}
\includegraphics[angle=0,trim=0in 0in 0in 0in, clip, width=0.75\textwidth]{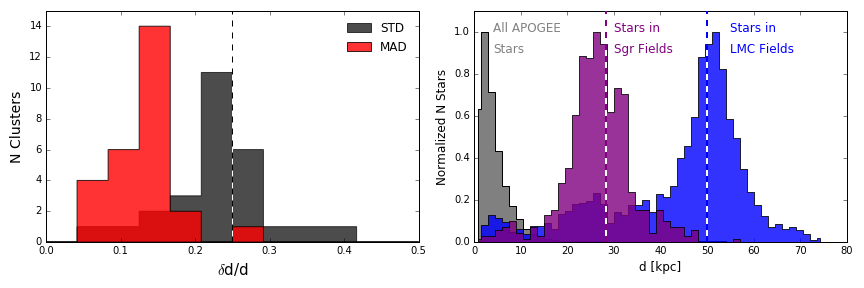} 
\caption{\textit{Left panel:} distance dispersion for stars in kinematically-identified open clusters (Poovelil et al.\ in prep), measured using the standard deviation (STD, black histogram) and the median absolute deviation (MAD, red histogram) of the cluster members' distance moduli. We adopt the more conservative typical uncertainty of 25\% (vertical dashed line) for our sample in this paper. \textit{Right panel:} distance distribution of stars targeted in fields towards the Large Magellanic Cloud (LMC) in blue and towards the Sagittarius (Sgr) dwarf galaxy in purple, compared with the distance distribution of the general APOGEE sample in gray. All three histograms have been scaled to peak at 1. The vertical dashed lines indicate the $d=28.2$~kpc distance to the Sgr core based on RRL \citep{Hernitschek_2019_RRLto_dSPhs} and the $d=49.88$~kpc distance to the LMC based on eclipsing binaries \citep{Pietrzynski_2013_LMCdistance}.}
\label{fig:distances_metrics}
\end{center}
\end{figure*}

We perform a number of tests and comparisons to verify the quality of our derived spectrophotometric distances, which are summarized in Figures~ \ref{fig:distances_methods} and \ref{fig:distances_metrics}.

In the left panel of Figure~\ref{fig:distances_methods}, we show the comparison of our spectrophotometric distances with those from the catalog of \textit{Gaia}-based Bayesian distances of \citet[][hereafter CBJ]{Bailer-Jones18}. We cross-match our catalog for the APOGEE giant sample ($\log{g} \le 3.5$) with that of CBJ and select stars with distances smaller than 2.5~kpc and relative error in parallaxes $\sigma_\varpi/\varpi \le 0.1$. We find a median offset of $(d-d_{\rm CBJ})/d=3.4$\% (i.e., 85~pc at 2.5~kpc), with a standard deviation of 21.1\%.

The middle panel displays the comparison of our distances with those from the neural network {\sc astroNN} \citep[aNN][]{Leung19}, for giant stars within 10~kpc.  We find a median offset of $(d-d_{\rm aNN})/d=4.8$\%, with a standard deviation of 21.1\%.

In the right panel our distances are compared with those computed with the {\sc StarHorse} code \citep[][SH]{Queiroz18} for giant stars within 10~kpc of the Sun.
Here, we find even better agreement than with other two methods, with a median offset of $(d-d_{\rm SH})/d=2.1$\% and a standard deviation of 13.4\%.

Figure~\ref{fig:distances_metrics} shows distance properties for some different systems using our spectrophotometric distances.  In the left panel, we show the fractional distance {\it spread} of several open clusters whose members are selected using dynamics only, with no assumption of distance (e.g., from isochrone-fitting; Poovelil et al.\ in prep).  The black histogram is the standard deviation of each cluster members' distances, and the red histogram shows the median absolute deviation. We adopt the more conservative typical uncertainty of 25\% (vertical dashed line) for our sample in this paper.

In the right panel of Figure~\ref{fig:distances_metrics}, we show the distance distribution of the entire APOGEE sample (gray histogram, predominantly within $d<10$~kpc), compared to two dwarf galaxies.  Stars targeted in the Sagittarius (Sgr) dwarf galaxy are shown in purple, and stars targeted in the Large Magellanic Cloud (LMC) are shown in blue.  We see the expected disc and halo contamination, but the samples are dominated by stars at distances entirely consistent with previously determined literature values --- here, shown as $d_{\rm Sgr}=28.2$~kpc based on RR Lyrae \citep{Hernitschek_2019_RRLto_dSPhs} and  $d_{\rm LMC}=49.88$~kpc based on eclipsing binaries \citep{Pietrzynski_2013_LMCdistance}.  The width of these peaks is entirely consistent with the 25\% uncertainty established from the left panel.

\bsp	
\label{lastpage}
\end{document}